\documentclass[final,1p,times,twocolumn]{elsarticle}
\usepackage{amssymb}
\usepackage{amsmath}
\usepackage[colorinlistoftodos, textsize=small]{todonotes}
\usepackage{booktabs}
\usepackage{tabularx}
\usepackage{array}
\usepackage{xcolor}
\usepackage{url}
\definecolor{revisionblue}{RGB}{0,0,0}
\newcommand{\rev}[1]{{\color{revisionblue}#1}}
\journal{Fundamental Research}

\begin{document}

\begin{frontmatter}

%% Title, authors and addresses

%% use the tnoteref command within \title for footnotes;
%% use the tnotetext command for theassociated footnote;
%% use the fnref command within \author or \affiliation for footnotes;
%% use the fntext command for theassociated footnote;
%% use the corref command within \author for corresponding author footnotes;
%% use the cortext command for theassociated footnote;
%% use the ead command for the email address,
%% and the form \ead[url] for the home page:
%% \title{Title\tnoteref{label1}}
%% \tnotetext[label1]{}
%% \author{Name\corref{cor1}\fnref{label2}}
%% \ead{email address}
%% \ead[url]{home page}
%% \fntext[label2]{}
%% \cortext[cor1]{}
%% \affiliation{organization={},
%%             addressline={},
%%             city={},
%%             postcode={},
%%             state={},
%%             country={}}
%% \fntext[label3]{}

\title{Probing Triton's Space Environment and Internal Structure: An Integrated Detection-and-Interpretation Framework}

%% use optional labels to link authors explicitly to addresses:
%% \author[label1,label2]{}
%% \affiliation[label1]{organization={},
%%             addressline={},
%%             city={},
%%             postcode={},
%%             state={},
%%             country={}}
%%
%% \affiliation[label2]{organization={},
%%             addressline={},
%%             city={},
%%             postcode={},
%%             state={},
%%             country={}}

\author[aff1,aff2]{Jiansen He\corref{cor1}} %% Author name
\author[aff1,aff3]{Chuanpeng Hou} %% Author name
\author[aff1]{Haoen Xie}
\author[aff1]{Jiaqi Li}
\author[aff1]{Tianhang Chen}
\author[aff1]{Hong Zou}
\author[aff1]{Xuzhi Zhou}
\author[aff2]{Hui Li}
\author[aff1]{Yan Li}
\author[aff4,aff5]{Fuchuan Pang}
\author[aff5]{Bingkun Yu}
\author[aff5]{Hui Huang}
\author[aff6]{Tong Wang}

\cortext[cor1]{Corresponding author. \textit{Email address:} jshept@pku.edu.cn}

%% Author affiliation
\affiliation[aff1]{organization={School of Earth and Space Sciences, Peking University},%Department and Organization
            %addressline={}, 
            city={Beijing},
            %postcode={}, 
            %state={},
            country={China}}

%% Author affiliation
\affiliation[aff2]{organization={State Key Laboratory of Solar Activity and Space Weather, National Space Science Center, Chinese Academy of Sciences},%Department and Organization
            %addressline={}, 
            city={Beijing},
            %postcode={}, 
            %state={},
            country={China}}

%% Author affiliation
\affiliation[aff3]{organization={Institut für Physik und Astronomie, Universität Potsdam},%Department and Organization
            %addressline={}, 
            city={Potsdam},
            %postcode={}, 
            %state={},
            country={Germany}}
            
\affiliation[aff4]{organization={Lunar Exploration and Space Engineering Center (LESEC), China National Space Administration (CNSA)},%Department and Organization
            %addressline={}, 
            city={Beijing},
            %postcode={}, 
            %state={},
            country={China}}

\affiliation[aff5]{organization={Deep Space Exploration Laboratory (DSEL)},%Department and Organization
            %addressline={}, 
            city={Hefei},
            %postcode={}, 
            state={Anhui},
            country={China}}
\affiliation[aff6]{organization={Institute of Spacecraft System Engineering (ISSE), China Academy of Space Technology (CAST)},
            city={Beijing},
            country={China}}

%% Abstract
\begin{abstract}
%% Text of abstract

%% 【注释，jshept, 20260827】Triton, Neptune’s largest moon and one of the most distant icy moons in the Solar System, is a compelling target for investigating the habitability of ocean worlds. A subsurface ocean may exist beneath its ice shell, and its depth, thickness, composition, and electrical conductivity are key parameters for understanding Triton’s internal structure, geologic activity, and interaction with Neptune’s magnetosphere. We present an integrated observational and modeling framework that combines planetary interior modeling, induced magnetic field analysis, magnetohydrodynamic simulations, and seismic-wave studies to constrain these subsurface properties. Interior models provide physically consistent profiles of density, temperature, electrical conductivity, and seismic velocity, which are used to calculate seismic responses, while magnetohydrodynamic simulations characterize the space plasma environment and associated magnetic perturbations. By combining magnetic, plasma, and seismic measurements, this multidisciplinary approach can reduce degeneracies among ocean depth, thickness, and conductivity, enabling more robust constraints on Triton’s subsurface ocean and improving assessment of its potential habitability. Such coordinated measurements should therefore be considered an important component of future Triton exploration missions.

\rev{
Triton, Neptune's largest moon, is a prime ocean-world target: its retrograde orbit, young surface, cryovolcanism, and dense ionosphere suggest geological activity and a possible subsurface ocean. Constraining ocean thickness, composition, and conductivity is essential for habitability assessment, but magnetic induction alone cannot resolve the thickness--conductivity degeneracy, and magnetic perturbations from Triton's space currents can obscure the internal induction signal.
We present an integrated detection-and-interpretation concept linking four physically consistent calculations. Using `PlanetProfile', we construct a common radial interior structure (temperature, density, conductivity, seismic-wave speed). We then employ `MoonMag' to compute the degree-one magnetic-induction response from that conductivity profile at the synodic, rotational, and orbital periods. We perform a multi-fluid `SWMF' simulation with the induced dipole as the inner-boundary condition and develop a Coulomb-gauge Poisson reconstruction to isolate space-current magnetic fields. Finally, we develop the `TritonSeis' workflow—three-dimensional seismic forward modeling plus hierarchical travel-time inversion—to constrain the ice--ocean and ocean--rock interface depths.
We find that induction is substantially more sensitive to ocean conductivity than to layer thickness, and that space-current fields are comparable in amplitude to the internal induction signal. A five-station synthetic recovery test resolves both interfaces to first order, with errors of +8.4\% for the ice shell and -12.5\% for the ocean. Under a conservative noise assumption, the minimum detectable magnitudes are approximately 3.8–4.6 at epicentral distances of 100–1000 km. The Poisson reconstruction and end-to-end seismic recovery are, to our knowledge, the first such quantitative demonstrations for Triton. Coordinated magnetic, plasma, and seismic measurements are complementary and can break the conductivity--thickness degeneracy, providing a framework for future Triton exploration.
}

\end{abstract}

%%Graphical abstract
%\begin{graphicalabstract}
%\includegraphics{grabs}
%\end{graphicalabstract}

%%Research highlights
%\begin{highlights}
%\item Research highlight 1
%\item Research highlight 2
%\end{highlights}

%% Keywords
%\begin{keyword}
%% keywords here, in the form: keyword \sep keyword

%% PACS codes here, in the form: \PACS code \sep code

%% MSC codes here, in the form: \MSC code \sep code
%% or \MSC[2008] code \sep code (2000 is the default)

%\end{keyword}

\end{frontmatter}

%% Add \usepackage{lineno} before \begin{document} and uncomment 
%% following line to enable line numbers
%% \linenumbers

%% main text
%%

\section{Introduction}

%%【注释】 Icy moons are a class of celestial bodies with significant research value in the solar system. Their interiors typically display layered structures—surface ice shell, liquid-water ocean, and possibly a rocky core—formed through long-term differentiation and accumulation. The study of subglacial oceans is crucial for exploring extraterrestrial life and habitable environments, as liquid water provides the basis for life\citep{chyba2001possible}. Interactions between such oceans and mineral-rich seafloors can generate organics and minerals through hydrothermal reactions\citep{hand2007energy}, while internal energy sources such as tidal heating sustain liquid oceans and drive large-scale circulation of energy and materials\citep{hussmann2006subsurface}. Ice shells also shield both the underlying oceans and any potential subsurface life from harmful radiation and impact damage\citep{cooper2001energetic}.

Icy moons are an important class of planetary bodies in the Solar System because many are thought to harbor subsurface liquid-water oceans beneath their outer ice shells. Their interiors are commonly characterized by layered structures consisting of an ice shell, a subsurface ocean, and a deeper rocky or differentiated interior formed through thermal and compositional evolution. Subsurface oceans are of particular interest in the search for extraterrestrial habitability because liquid water is a fundamental requirement for life as we know it \citep{chyba2001possible}. Chemical interactions between ocean water and mineral-rich rocky interiors may generate reduced compounds, minerals, and organic species through hydrothermal processes \citep{hand2007energy}, potentially providing chemical energy for biological activity. Internal heat sources, including tidal dissipation and radiogenic heating, can help maintain subsurface oceans over geological timescales and drive the transport of heat and materials within these bodies \citep{hussmann2006subsurface}. In addition, overlying ice shells provide physical shielding from energetic radiation and impact processes, helping to preserve relatively stable subsurface environments \citep{cooper2001energetic}.

%%【注释】 Key geophysical parameters—ice shell thickness, ocean depth, solute composition, and electrical conductivity—are essential for understanding internal structures, thermal evolution, and life-support potential. These can be constrained by theoretical modeling, orbital measurements, and in-situ exploration. Models predict internal processes and guide observations; spacecraft measurements of gravity and magnetic fields enable inversion for ocean properties, as demonstrated for Europa\citep{khurana1998induced}; and in-situ surface or subsurface investigations can directly reveal structural, compositional, and thermal characteristics of Titan \citep{lorenz2019titan}. Ice shell activity and volatile release further influence atmospheric and ionospheric formation, where gases are ionized by solar radiation and particle impacts to form ionospheres\citep{tyler1989voyager, kliore1997ionosphere, buccino2021dual}. However, key questions remain about ion composition, energy balance, and magnetospheric coupling.

Constraining key geophysical properties, including ice-shell thickness, ocean depth and thickness, solute composition, and electrical conductivity, is therefore essential for understanding the internal structure, thermal evolution, and habitability of icy moons. These properties can be investigated through a combination of theoretical modeling, remote spacecraft observations, and in situ measurements. Interior models provide physically consistent predictions of internal structure and evolution, while measurements of gravity and magnetic fields can be used to infer subsurface properties; for example, magnetic induction observations have provided strong evidence for a conductive subsurface ocean at Europa \citep{khurana1998induced}. In situ surface or subsurface investigations can provide additional constraints on local structure, composition, and thermal conditions, as considered for future exploration of Titan \citep{lorenz2019titan}. Surface and subsurface activity, including volatile release through plumes or cryovolcanism, can also modify the surrounding atmosphere and plasma environment. Neutral gases released from icy moons may be ionized by solar radiation and energetic particle impacts, contributing to the formation and variability of their ionospheres \citep{tyler1989voyager,kliore1997ionosphere,buccino2021dual}. Understanding these coupled processes remains important for determining ionospheric composition and energetics, as well as the interaction between icy moons and their surrounding magnetospheric environments.

%%【注释】 Europa and Enceladus have yielded notable results. Pioneer and Voyager flybys initiated Europa studies in the 1970s, followed by Galileo’s 1995–2003 observations, which inferred ocean depths of tens to hundreds of kilometers and ice shells several to tens of kilometers thick\citep{kivelson1999europa, kivelson2000galileo, zimmer2000subsurface}. Models suggest a saline ocean with measurable conductivity\citep{vance2021magnetic}. Similarly, Enceladus, first noted for unusual surface features by Voyager\citep{smith1981encounter}, was later confirmed by Cassini (2004–2017) to host an active subsurface ocean, ejecting plumes containing water, ammonia, and organics\citep{porco2006cassini, waite2017cassini}, confirming an active subsurface ocean.

Europa and Enceladus have provided important evidence for subsurface oceans and demonstrated the scientific value of icy-moon exploration. Galileo observations of Europa strongly supported the presence of a global subsurface ocean beneath an ice shell several to tens of kilometers thick \citep{kivelson1999europa,kivelson2000galileo,zimmer2000subsurface}, while models suggest that the ocean is saline and electrically conductive \citep{vance2021magnetic}. At Enceladus, Cassini detected active plumes containing water, ammonia, and organic compounds \citep{smith1981encounter,porco2006cassini,waite2017cassini}, providing compelling evidence for an active subsurface ocean.

%%【注释】 Triton, Neptune’s largest moon, is distinctive for its retrograde orbit, cryovolcanism, sub-ice ocean, and dense ionosphere, making it a prime target for future exploration\citep{blanc2021science}. Voyager 2’s 1989 flyby revealed young terrains, ongoing geological activity, and a nitrogen-rich atmosphere with trace methane and CO\citep{smith1989voyager, tyler1989voyager}, all suggesting a warm subsurface ocean beneath its frigid surface. Comprehensive exploration will require multiple flybys, low-altitude orbiters, surface missions, and potentially drilling or submersible probes. Modeling Triton’s space environment, magnetospheric coupling, and internal ocean will guide mission design. While electromagnetic measurements are essential for probing ocean conductivity and thickness, they suffer from a trade-off between the two. Seismology can provide independent constraints on thickness, helping to resolve this ambiguity and enabling a more robust characterization of the subsurface ocean. 

Triton, Neptune’s largest moon, is distinguished by its retrograde orbit, young surface, cryovolcanic activity, atmosphere, and ionosphere, making it an important target for future exploration \citep{blanc2021science}. Voyager 2 revealed active plumes, geologically young terrains, and a nitrogen-dominated atmosphere with trace methane and CO \citep{smith1989voyager,tyler1989voyager}, supporting the possibility of a present-day subsurface ocean. Characterizing this ocean requires constraints on its depth, thickness, composition, and electrical conductivity, together with an understanding of Triton’s interaction with Neptune’s magnetosphere. Magnetic induction can probe a conductive ocean, but its response is often degenerate with respect to ocean thickness and conductivity. Seismic measurements provide independent constraints on layer thicknesses and can help resolve this ambiguity.

%%【注释】 \rev{To examine this trade-off within a common physical framework, the calculations are linked through their inputs and observables. PlanetProfile provides the radial thermodynamic, electrical, and elastic structure. MoonMag uses the conductivity profile to calculate the periodic induced response. The multi-fluid SWMF calculation evaluates the magnetic contribution of plasma currents generated by the interaction between Neptune's magnetosphere and Triton's atmosphere and ionosphere. Seismic travel times sample the interfaces represented by the same radial interior model and supply an independent constraint on layer thickness. The resulting magnetic, plasma, and seismic observables therefore address complementary parts of the subsurface-ocean problem.}

To address these complementary constraints, we combine interior, electromagnetic, plasma, and seismic models within a common framework. PlanetProfile, an open-source interior-structure modeling code \citep{styczinski2023planetprofile} provides the radial thermodynamic, electrical, and elastic structure. MoonMag, a spherical harmonic magnetic-induction solver for layered conductors \citep{styczinski2022perturbation}, is used to calculate the induced magnetic response to the internal structure. Multi-fluid Space Weather Modeling Framework (SWMF) simulations \citep{Toth2005SWMF,rubin2015self} quantify space magnetic fields generated by Triton’s interaction with Neptune’s magnetosphere, while seismic travel-time calculations provide independent constraints on the ice-shell and ocean thicknesses. Together, these magnetic, plasma, and seismic observables enable a more robust characterization of Triton’s subsurface structure.

%%【注释】 This study focuses on Triton, integrating geophysics and space physics within a coupled-spheres framework. Using physical modeling, we simulate Triton’s interior and space environment to develop exploration strategies. Section 2 models the internal structure under observational constraints, including conductivity \rev{and seismic-velocity} profiles. Section 3 examines induced magnetic fields from Neptune’s rotation and Triton’s orbital dynamics. Section 4 simulates the coupling between Neptune’s magnetosphere and Triton’s ionosphere and its effect on ambient fields. Section 5 introduces a seismic inversion approach that, when combined with magnetic and current measurements, enables robust joint determination of Triton’s ice shell and ocean thickness. Section 6 reflects on theoretical modeling and detection methodologies, offering recommendations for payload design and mission planning. Section 7 summarizes the key findings and outlines future directions for advancing Triton and icy moon exploration.

This study integrates geophysical and space-physics modeling to characterize Triton’s interior and surrounding plasma environment and to assess measurement strategies for future exploration. Section 2 presents the interior structure, conductivity, and seismic-velocity profiles. Section 3 examines magnetic induction, Section 4 investigates magnetospheric interaction and associated magnetic perturbations, and Section 5 develops seismic constraints on the ice shell and ocean. Section 6 discusses implications for observational strategies and mission design, and Section 7 summarizes the main conclusions.

\section{Internal Stratified Structure and Conductivity Distribution of Icy Moon}

\subsection{Model Principles}

We assume that Triton's interior consists of an ice shell, subsurface ocean, and rock layer in sequence. \rev{Depending on the pressure--temperature path, the hydrosphere may also contain high-pressure ice phases, whose stability is evaluated from the corresponding equation of state.} For the pressure profile of Triton, the surface pressure is 0; at the ice--ocean interface, the pressure is obtained from the preset temperature and the equation of state (EOS) for ice Ih. \rev{PlanetProfile \citep{styczinski2023planetprofile} then integrates the hydrostatic pressure inward while adjusting the layer radii to satisfy the prescribed bulk constraints. For the interior models with MgSO$_4$ ocean salinities of 1 and 10~g~kg$^{-1}$, the ocean–rock interface is determined by searching up to an upper pressure bound of 250~MPa. The 10~g~kg$^{-1}$ MgSO$_4$ model converges at 201~MPa, well below the upper search limit. High-pressure ice is diagnosed along the integrated pressure--temperature trajectory. At each hydrosphere grid point, the calculated pressure, temperature, and composition are queried against the phase table; a high-pressure-ice layer is diagnosed if the trajectory enters a stable ice-II, ice-III, ice-V, or ice-VI field before reaching the silicate interface.}

%%【注释，jshept, 20260828】There are two \rev{heat-transport regimes} for calculating the internal temperature profile of Triton: one is through heat conduction, where temperature is a function of pressure; the other considers thermal convection, where according to the adiabatic assumption, $\frac{\partial T}{\partial P} = \frac{\alpha T}{\rho C_{p}}$, so temperature is also a function of pressure. For the surface ice layer of Triton, the Rayleigh number (Ra) is calculated to determine the convection status, and then the corresponding equation is selected to calculate the temperature profile. \rev{The 10~g~kg$^{-1}$ reference solution has a Rayleigh number of $Ra=9.28\times10^{7}$, exceeding the critical value of $Ra_{\rm crit}=7.18\times10^{6}$ for the surface ice layer. This results in a stagnant-lid structure consisting of a 74.67 km conductive lid, a 36.74 km convecting layer, and a 1.92 km basal thermal boundary layer.} The subsurface ocean utilizes the adiabatic equation of convection to calculate its temperature profile, while the rock layer employs the heat conduction equation.

\rev{The temperature profile is assigned according to the heat-transport regime appropriate for each layer. In the ice shell, PlanetProfile evaluates the Rayleigh number,
\[
Ra=
\frac{\alpha C_P\rho^2 g\Delta T_{\rm ice}h_{\rm ice}^{\,3}}
     {\eta_{\rm conv}k},
\]
where \(\alpha\) is the thermal expansivity, \(C_P\) is the isobaric specific heat, \(\rho\) is density, \(g\) is gravitational acceleration, \(\Delta T_{\rm ice}\) is the temperature contrast across the ice shell, \(h_{\rm ice}\) is the ice-shell thickness, \(\eta_{\rm conv}\) is the viscosity of the convecting ice, and \(k\) is thermal conductivity. The calculated value is compared with the critical Rayleigh number \(Ra_{\rm crit}\) for the adopted stagnant-lid parameterization. For the 10~g~kg\(^{-1}\) MgSO$_4$ reference solution, \(Ra=9.28\times10^7\), which exceeds \(Ra_{\rm crit}=7.18\times10^6\). The ice shell is therefore predicted to undergo solid-state convection in part of its interior and is represented by a stagnant-lid structure with a 74.67~km conductive lid, a 36.74~km convecting layer, and a 1.92~km basal thermal boundary layer.}

\rev{The liquid ocean is treated as a convecting, well-mixed layer and follows an adiabatic temperature gradient. The silicate interior is treated with spherically symmetric steady conductive heat transport. In the reference model, the silicate radiogenic-heating rate is \(4.5\times10^{-12}\)~W~kg\(^{-1}\), and silicate tidal heating is set to zero.}

\subsection{Model Setup}

%%【注释】 We set the following parameters for each layer inside Triton: the surface temperature is 38 K, and the surface pressure is set to zero. At the interface between the ice shell and the subsurface ocean, the temperature is set to 266 K. The solute in the subsurface ocean is magnesium sulfate ($MgSO_4$)\citep{styczinski2022perturbation}, considering two salinity conditions: low salinity (1 g~kg$^{-1}$) and high salinity (10 g~kg$^{-1}$). \rev{MgSO$_4$ provides a reference composition for which PlanetProfile supplies internally consistent thermodynamic and electrical properties. A 10~g~kg$^{-1}$ PlanetProfile Seawater endmember is also calculated to examine composition sensitivity; this option represents mixed seawater through the Gibbs Seawater/TEOS-10 formulation. The silicate radiogenic-heating rate is $4.5\times10^{-12}$~W~kg$^{-1}$ and the silicate tidal-heating term is set to zero in the reference calculation.}

PlanetProfile package \citep{styczinski2023planetprofile} is used to estimate the internal physical properties, electromagnetic characteristics, and seismological features of Triton. The surface temperature is set to 38 K and the surface pressure to zero, while the temperature at the ice-ocean interface is fixed at 266 K. The subsurface ocean is assumed to contain magnesium sulfate (MgSO$_4$) \citep{styczinski2022perturbation}, with two salinities of 1 and 10~g~kg$^{-1}$. MgSO$_4$ is adopted as a reference composition because PlanetProfile provides internally consistent thermodynamic and electrical properties for this solution. To assess compositional sensitivity, we also consider a 10~g~kg$^{-1}$ Seawater ocean based on the Gibbs Seawater/TEOS-10 formulation. The silicate radiogenic-heating rate is set to $4.5\times10^{-12}$~W~kg$^{-1}$, while silicate tidal heating is neglected. 
Initially, the internal stratification of the icy moon is assumed, such as an ice layer, a subsurface ocean, and a rock layer from the outside in. \rev{Pure-H$_2$O ice-Ih and high-pressure-ice properties are evaluated with SeaFreeze and the Gibbs-energy representation of water-ice phases \citep{journaux2020holistic}. The saline reference ocean uses PlanetProfile's MgSO$_4$ thermodynamic tables based on the treatment of Vance and Brown \citep{vance2013thermodynamic}, and the silicate properties are obtained from Perple\_X-based tables. The development of a metallic core depends on differentiation, oxidation state, composition, and thermal history. We therefore retain a core-free structure with a 10~g~kg$^{-1}$ MgSO$_4$ ocean as the reference model and calculate a separate iron-core case as a sensitivity test.}
%% 【注释】 added “with a 10~g~kg$^{-1}$ MgSO$_4$ ocean” by 谢浩恩，2026.8.23

\subsection{Modeling Results}

\begin{figure}[h!]%
\centering
\includegraphics[width=1.0\textwidth]{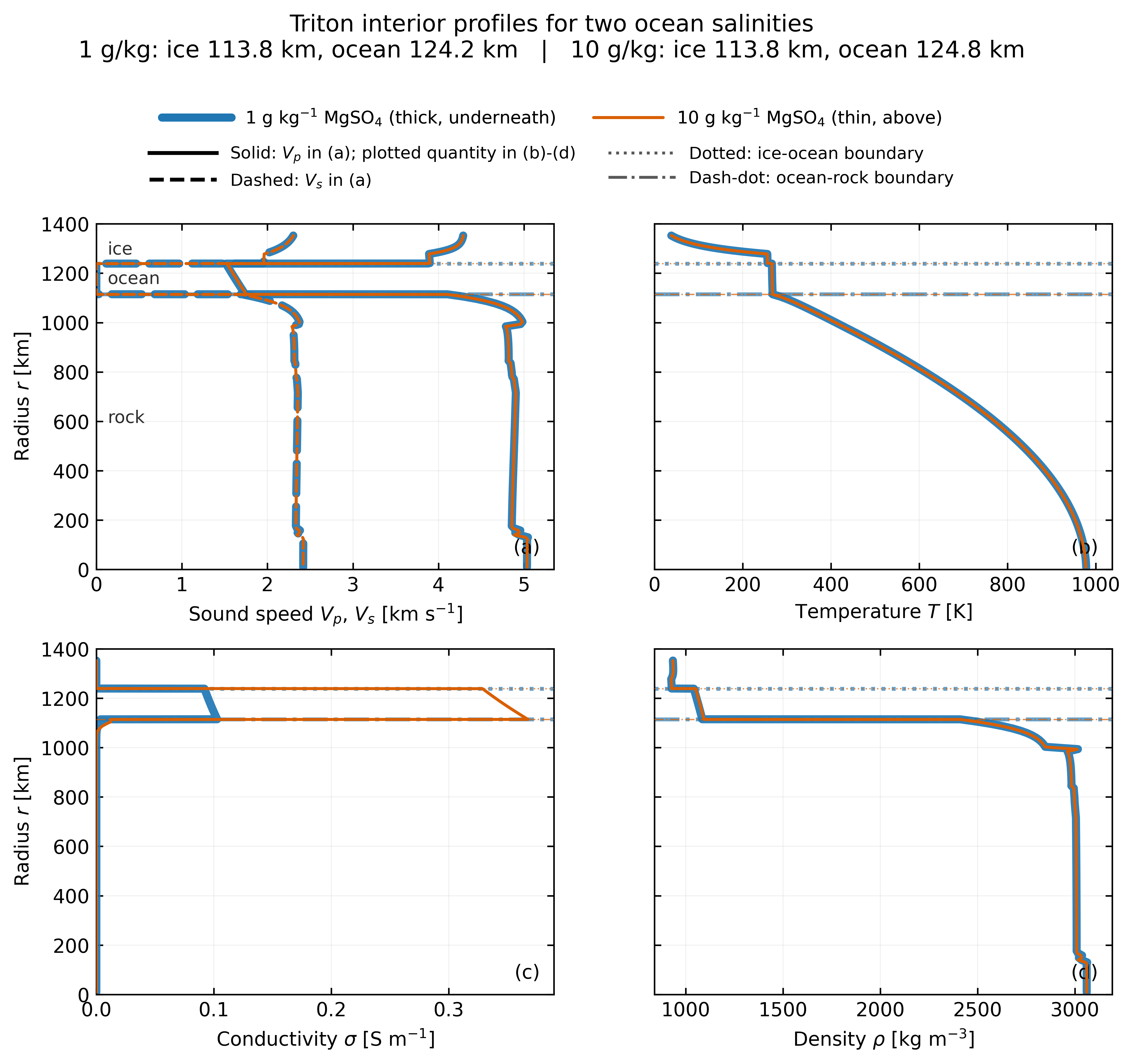}
\caption{\rev{Profile distributions of Triton parameters computed with PlanetProfile for oceans containing 1 and 10~g~kg$^{-1}$ MgSO$_4$. Thick blue and thin orange curves show the two salinities of 1 and 10~g~kg$^{-1}$ MgSO$_4$, respectively. (a) P-wave velocity (solid) and S-wave velocity (dashed); (b) temperature; (c) electrical conductivity; and (d) density. Dotted and dash-dotted horizontal lines mark the ice-ocean and ocean-rock interfaces.}
}\label{fig1}
\end{figure}

Based on the above model settings, we obtain the radial profiles of Triton interior (Figure~\ref{fig1}), including the wave propagation speed of longitudinal waves (P-waves) and transverse waves (S-waves), temperature, conductivity, and density. The main structural parameters of the reference and sensitivity models are summarized in Table~\ref{tab:interior_sensitivity}. According to the model calculation, \rev{the 10~g~kg$^{-1}$ MgSO$_4$ reference structure has an ice shell 113.75~km thick, and a liquid ocean approximately 124.80~km thick above the internal rock layer. The ice-ocean interface is at 85.294~MPa and 266~K, while the ocean--rock interface is at approximately 201.3~MPa and 267.65~K. The relatively low seismic velocities at the top of the silicate layer reflect the porosity adopted in the model; the higher elastic moduli at greater depth produce P-wave velocities above those in the ice shell.} By comparing different MgSO$_4$ salinity of \rev{1 and 10~g~kg$^{-1}$ in Figure~\ref{fig1}}, we find that, as expected, the salinity of the subsurface ocean significantly alters the conductivity of the ocean layer, while having only a minor effect on the thickness of the ice shell. When the $MgSO_4$ concentration is 1 g~kg$^{-1}$, the conductivity $\sigma$ is approximately 0.1 S~m$^{-1}$; when the $MgSO_4$ concentration is 10 g~kg$^{-1}$, the conductivity $\sigma$ is in the range of [0.3, 0.4] S~m$^{-1}$. Therefore, if the conductivity of the ocean layer can be determined from measurements of the induced magnetic field, as described in the following section, the salinity of \rev{a specified ocean composition and thermal state} can then be conditionally inferred.
%% 【注释】 added “The main structural parameters of the reference and sensitivity models are summarized in Table~\ref{tab:interior_sensitivity}.” by 谢浩恩，2026.8.23

\begin{figure}[htbp]%
\centering
\includegraphics[width=1.0\textwidth]{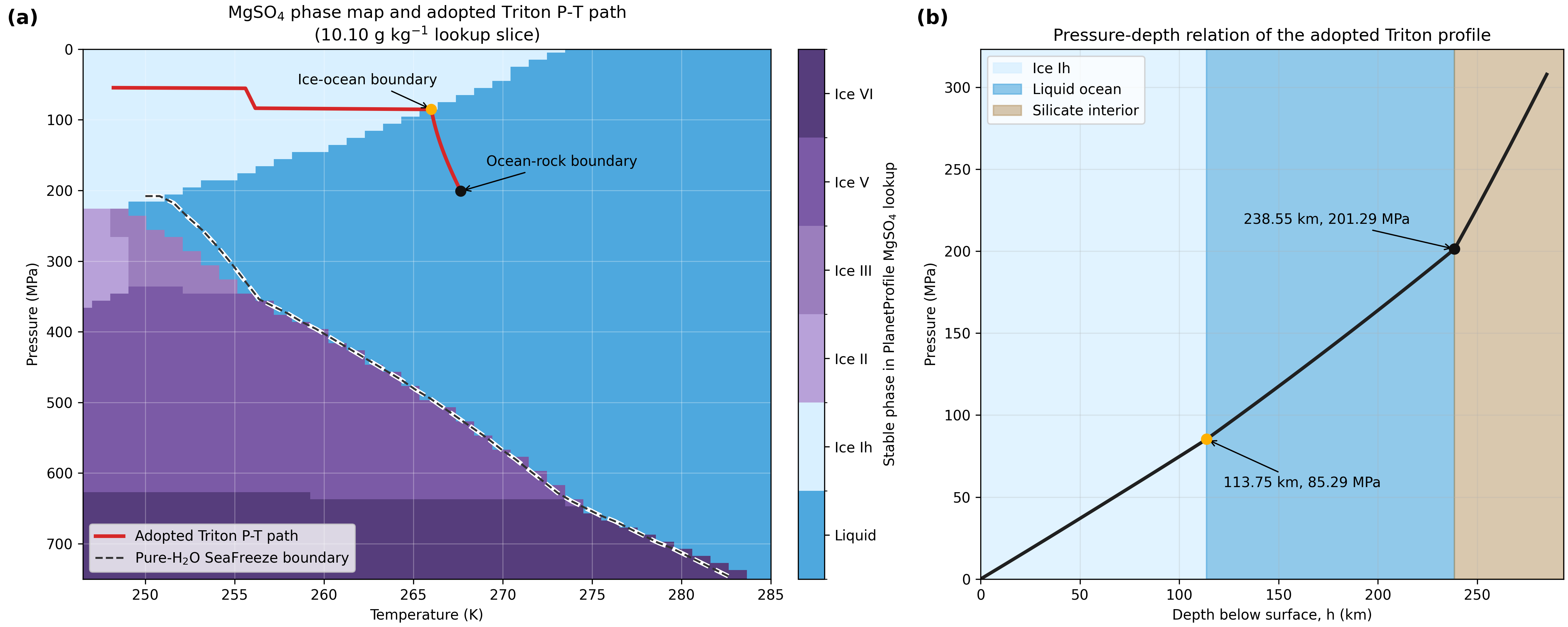}
\caption{\rev{High-pressure-ice diagnostic for the 10~g~kg$^{-1}$ MgSO$_4$ reference model. (a) The PlanetProfile hydrosphere pressure--temperature trajectory superposed on the nearest available MgSO$_4$ phase-lookup slice (10.10~g~kg$^{-1}$). The trajectory enters the liquid regime at the ice--ocean boundary and reaches the ocean floor at approximately 201.3~MPa and 267.65~K. At the nearest tabulated temperature, ice V begins at approximately 531.8~MPa. The dashed curve gives the pure-H$_2$O SeaFreeze liquid--high-pressure-ice boundary as an independent comparison. (b) Pressure versus depth for the same reference profile. The ice--ocean and ocean--rock interfaces occur at 113.75 and 238.55~km, respectively, and the pressure increase below the latter interface takes place within the silicate interior. Together, the two panels show that the adopted hydrosphere terminates within the liquid stability regime, the seafloor pressure is approximately 330.7 MPa lower than the MgSO$_4$ ice-V onset pressure}
}\label{fig2}
\end{figure}

\rev{Figure~\ref{fig2} evaluates high-pressure-ice stability along the modeled pressure–temperature trajectory and within the corresponding radial structure. In Figure~\ref{fig2}a, the trajectory passes through the ice-Ih regime, reaches the ice–ocean boundary at 85.294 MPa and 266 K, and remains within the liquid regime to the base of the ocean. The last liquid grid point occurs at 201.044 MPa and 267.648 K, followed by the first silicate grid point at 201.294 MPa and 267.653 K. On the nearest temperature slice of the $10.10~g~kg^{-1} MgSO_4$ lookup, 267.88 K, the first high-pressure solid encountered with increasing pressure is ice V at 531.773 MPa. The reference ocean therefore terminates approximately 330.7 MPa before this phase transition. The pure-$H_2O$ SeaFreeze calculation gives an ice-V onset near 524 MPa at the seafloor temperature. Figure~\ref{fig2}b shows that the hydrosphere ends at the ocean–rock interface at 238.55 km depth, below which pressure increases within the silicate interior. The modeled hydrosphere consequently terminates while its pressure–temperature trajectory remains in the liquid stability regime, allowing direct contact between the ocean and the porous silicate interior. This diagnosis applies to the adopted bulk constraints, thermal profile, and ocean composition.}

\rev{Two additional calculations illustrate the sensitivity to composition and differentiation (Table~\ref{tab:interior_sensitivity}). At the same nominal concentration, the Seawater endmember has a mean ocean conductivity of 0.885~S~m$^{-1}$, about 2.54 times the MgSO$_4$ reference value, together with modest shifts in the ice and ocean thicknesses. The iron-core sensitivity case retains a 10~g~kg$^{-1}$ MgSO$_4$ ocean and contains a metallic core with a radius of 483.74~km; its ocean thickness increases to 163.86~km, while the silicate layer between the ocean and the metallic core is 591.25~km thick. Its mean ocean conductivity is 0.355~S~m$^{-1}$, close to the 0.348~S~m$^{-1}$ value of the core-free MgSO$_4$ reference model.}

\begin{table*}[htbp]
\centering
%%【注释，jshept, 20260828】\caption{Reference and sensitivity interior models of Triton. The silicate-layer thickness extends from the ocean--rock interface to the center in the core-free models and to the silicate--core boundary in the iron-core model. A dash indicates that no metallic core is included.}
\caption{Reference and sensitivity interior models of Triton. The
silicate-layer thickness is measured from the ocean-rock interface to
Triton's center in the core-free models and from the ocean-rock interface to
the silicate-core boundary in the iron-core model. A dash indicates that no
metallic core is included.}
\label{tab:interior_sensitivity}
\small
\setlength{\tabcolsep}{4pt}
\renewcommand{\arraystretch}{1.2}
\begin{tabularx}{\textwidth}{
    >{\raggedright\arraybackslash}p{0.10\textwidth}
    >{\raggedright\arraybackslash}p{0.19\textwidth}
    *{5}{>{\centering\arraybackslash}X}
}
\toprule
Case &
Ocean composition &
\shortstack{Ice-shell\\thickness (km)} &
\shortstack{Ocean\\thickness (km)} &
\shortstack{Silicate-layer\\thickness (km)} &
\shortstack{$\langle\sigma_{\rm oc}\rangle$\\(S~m$^{-1}$)} &
\shortstack{Metallic-core\\radius (km)} \\
\midrule

MgSO$_4$ reference &
MgSO$_4$, 10~g~kg$^{-1}$ &
113.75 &
124.80 &
1114.05 &
0.348 &
-- \\

Seawater endmember &
Seawater, 10~g~kg$^{-1}$ &
104.11 &
134.97 &
1113.52 &
0.885 &
-- \\

Iron-core endmember &
MgSO$_4$, 10~g~kg$^{-1}$ &
113.75 &
163.86 &
591.25 &
0.355 &
483.74 \\

\bottomrule
\end{tabularx}
\end{table*}
%% 【modified by 谢浩恩 2026.8.23】

\section{Inductive Response of Triton to External Changing Magnetic Fields in the Neptune System}

\subsection{Model Principles}

After obtaining the radial profile of electrical conductivity within Triton, we can calculate its response to external changing magnetic fields and the resulting induced magnetic fields in the Neptune system. The variations in external magnetic fields primarily stem from the rotation of the ice giant planet with its intrinsic magnetic fields and the revolution of the icy moon around the ice giant planet, along with magnetic field changes associated with the dynamics of the ice giant planetary magnetosphere.

Following the magnetic-induction formulation implemented in MoonMag \citep{styczinski2022perturbation}, the magnetic field within each conductive layer of a spherically symmetric body subjected to a time-varying external magnetic field obeys the magnetic diffusion equation
\begin{equation}
\nabla^2 \mathbf{B} = \mu \sigma \frac{\partial \mathbf{B}}{\partial t}
\label{eq:magnetic_diffusion}
\end{equation}
where $\mu$ is the magnetic permeability and $\sigma$ is the electrical conductivity, both of which vary with spatial position. Outside the icy moon, since the conductivity of the ionosphere is much lower than that of the internal ocean, both the external and induced magnetic fields approximately satisfy Laplace's equation $\nabla^2 \mathbf{B} = 0$. Under this approximation, the net magnetic field outside the icy moon can be expressed as:
\begin{equation}
\mathbf{B}_{\text{net}} = -\nabla \left[ \sum_{n,m} R \left( B_{nm}^e \left( \frac{r}{R} \right)^n + B_{nm}^i \left( \frac{R}{r} \right)^{n+1} \right) Y_{nm}(\theta, \phi) \right]
\label{eq:net_B_outside_icy_moon}
\end{equation}
where \(B_{nm}^e\) and \(B_{nm}^i\) are the external and induced magnetic-field coefficients, respectively, and \(Y_{nm}\) is the spherical-harmonic function. \rev{The reference radius $R$ is Triton's mean solid-surface radius, $r$ is the distance from its center, and the factor $(R/r)^{n+1}$ gives the radial decay of the induced contribution outside Triton.} Solving the magnetic diffusion equation in each radial layer and enforcing the electromagnetic interface conditions gives the complex induction response \citep{seufert2011multi,styczinski2022perturbation},
\begin{equation}
\frac{B_{nm}^i}{B_{nm}^e} = \frac{n}{n+1} A e^{i\phi}
\label{eq:ratio_Bi_to_Be}
\end{equation}
where $A$ represents the amplitude ratio of the induced magnetic field to the external magnetic field, and $\phi$ is the phase difference between the two fields. \rev{The complex response $A e^{i\phi}$ is obtained by solving the magnetic diffusion equation in every conductive layer and matching the magnetic-field components at each interface and at the surface. Equation~\ref{eq:ratio_Bi_to_Be} summarizes the response of the complete layered conductor.} 
\rev{In the calculations presented here, the imposed external magnetic field is represented only by its degree-one component. Because the adopted conductivity structure is spherically symmetric, different spherical-harmonic degrees do not couple, and a degree-one external excitation produces only the corresponding degree-one induced response. Therefore, \(n=1\) in Eq.~\ref{eq:ratio_Bi_to_Be}, and the calculated response is an induced dipole. The detailed calculation of \(A e^{i\phi}\) is given in Eqs.~21-23 of Seufert et al.~(2011) \citep{seufert2011multi}.}

\subsection{Model Setup}

Triton orbits Neptune at a distance of approximately 14.3 Neptune radii with an orbital period of about 141 h, whereas Neptune rotates with a period of about 16 h. Because Triton follows a retrograde orbit opposite to Neptune’s rotation, the corresponding synodic period is approximately 14 h. As Triton moves through Neptune’s magnetosphere, it is therefore exposed to time-varying magnetic fields associated primarily with the orbital, rotational, and synodic periods. These periodic variations are used to calculate the magnetic induction response and the resulting total magnetic field around Triton.

The induced magnetic response depends strongly on Triton’s internal conductivity structure. The thickness and composition of the ice shell, subsurface ocean, and rocky interior determine the radial conductivity profile, with the conductive subsurface ocean expected to provide the dominant contribution because of its relatively high conductivity and shallow depth. To examine this dependence, we consider ocean thicknesses of 0–200 km and electrical conductivities of 0.1–100 S m$^{-1}$ for an MgSO$_4$ ocean, and calculate the amplitude and phase of the induced magnetic field over this parameter space.

During the Voyager 2 encounter with the Neptune system, radio occultation measurements provided constraints on Triton’s ionosphere. Inversion of the dual-frequency phase delay yielded an electron density profile with a peak density of approximately $(20$–$50)\times10^{3}$ cm$^{-3}$ near an altitude of 350 km \citep{tyler1989voyager}. To account for the ionospheric contribution to the electromagnetic response, we include a conductive layer with $\sigma=0.05$ S m$^{-1}$ between altitudes of 250 and 450 km above Triton’s surface \citep{styczinski2022perturbation}.

%% 【注释，20260826】We use the \rev{MoonMag} package\citep{styczinski2022perturbation} to calculate the induced magnetic moments within Triton. \rev{For each external period, the radial conductivity profile is supplied to the layered spherical solver, which returns the complex degree-one response. This step converts the interior conductivity structure into the amplitude and phase of the magnetic field that can be measured outside Triton.}

We use MoonMag \citep{styczinski2022perturbation}, specifically its layered, spherically symmetric induction solver, to calculate Triton's induced magnetic response. For each of the synodic, rotational, and orbital forcing periods, the radial locations and electrical conductivities of the concentric layers are supplied to the solver, including the conductive ionospheric layer when specified. MoonMag solves the magnetic-diffusion problem in the radial layers and returns the complex degree-one induced-field coefficients, from which the amplitude and phase of the induced dipole field at Triton's surface are calculated. Although MoonMag can also represent nonspherical boundaries, no nonspherical boundary terms are included in the results presented here.

\subsection{Simulation Results}

\begin{figure}[htbp]%
\centering
\includegraphics[width=1.0\textwidth]{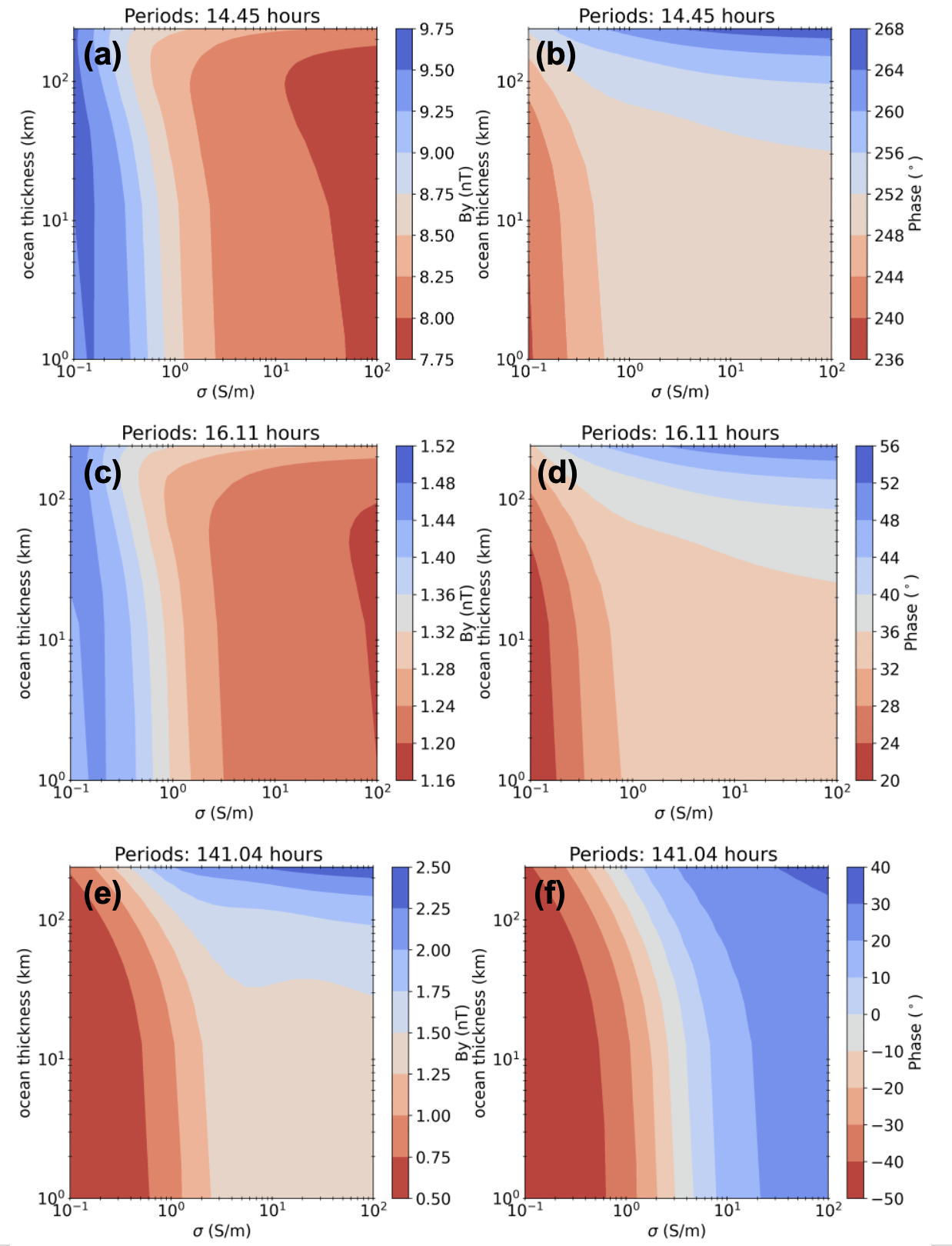}
\caption{
%% 【注释，20260826】Distributions of induced magnetic field intensity and phase across the parameter space of the electrical conductivity and thickness of the subsurface ocean, under external magnetic field environments with varying periods. $B_y$ denotes the maximum amplitude of the induced magnetic field component at Triton’s surface along the direction pointing toward Neptune. The calculation of the induced magnetic field includes the contribution from Triton’s ionosphere.
\rev{Amplitude and phase of the degree-one induced magnetic response across the parameter space of subsurface-ocean electrical conductivity and thickness. From top to bottom, the rows correspond to the Triton--Neptune synodic period (14.45 h), Neptune's rotation period (16.11 h), and Triton's orbital period (141.04 h), respectively. The left column (a, c, e) shows the maximum amplitude of \(B_y\) at Triton's surface, where the \(y\)-direction points toward Neptune, and the right column (b, d, f) shows the corresponding phase. The calculation includes Triton's conductive ionospheric layer.}
}\label{fig3}
\end{figure}

Figure~\ref{fig3} illustrates the intensity and phase distribution of the $y$-component of the induced magnetic field under different ocean thicknesses, conductivities, and external periodic magnetic fields. The coordinate system used is the Triton-Phi-Orbital (TPhiO) system, where the $y$-axis points toward Neptune, the $z$-axis aligns with Triton's spin axis, and the $x$-axis is determined by the right-hand rule. Figure~\ref{fig3} shows that the intensity of the induced magnetic field is primarily controlled by variations in the external magnetic field, with a period of 14.45 hours, corresponding to the synodic period between Neptune and Triton. The induced responses to short-period (14.45 hours and 16.11 hours) and long-period (141.04 hours) magnetic field excitations exhibit significant differences. Thus, after obtaining actual magnetic field observation data in the future, it may be possible to infer the depth and conductivity of Triton's subsurface ocean by analyzing the intensities of magnetic field disturbances at different periods.

From the contour patterns in Figure~\ref{fig3}, we note that when the ocean thickness is within tens of kilometers, the intensities and phases of the induced magnetic fields at three different periods are more sensitive to changes in ocean conductivity than to changes in ocean thickness. This insensitivity poses challenges for determining ocean thickness through induced magnetic field information alone, necessitating the consideration of other combined methods.

\begin{figure}[htbp]%
\centering
\includegraphics[width=1.0\textwidth]{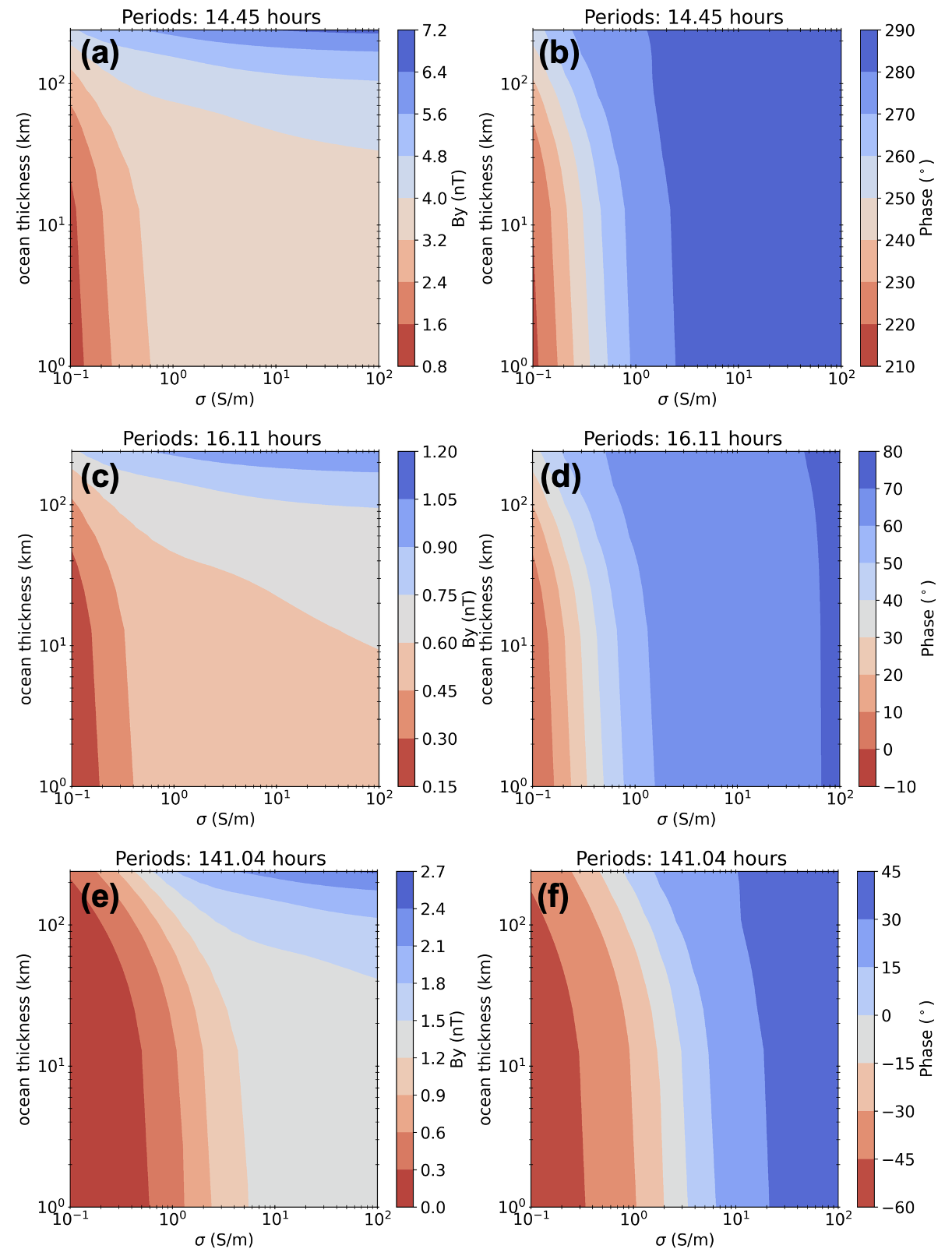}
\caption{
%%【注释】Distributions of induced magnetic field intensity and phase across the parameter space of the electrical conductivity and thickness of the subsurface ocean, under external magnetic field environments with varying periods. The layout of Figure 4 is consistent with that of Figure 3, except that the calculation of the induced magnetic field in Figure 4 excludes the influence of Triton’s ionosphere.
\rev{Same as Figure~\ref{fig3}, but excluding Triton's conductive ionospheric layer. From top to bottom, the rows correspond to the Triton--Neptune synodic period (14.45 h), Neptune's rotation period (16.11 h), and Triton's orbital period (141.04 h), respectively. The left column (a, c, e) shows the \(B_y\) amplitude, and the right column (b, d, f) shows the corresponding phase.}
}\label{fig4}
\end{figure}

When short-period variations exist in the external magnetic field, the intensity of the induced magnetic field decreases as the depth and conductivity of the subsurface ocean increase. This phenomenon primarily arises from the inclusion of the ionosphere in the model as an additional conductive layer, which significantly contributes to the induced magnetic field near Triton's surface. For comparison, we also calculate the intensity of the induced magnetic field without the ionosphere (Figure~\ref{fig4}), which shows a positive correlation between the induced magnetic field intensity and both ocean depth and conductivity, differing significantly from the results with the ionosphere (Figure~\ref{fig3}). \rev{The sensitivity of the induced‑field response to ionospheric inclusion is consistent with the flyby‑simulation results of Cochrane et al. (2022) \citep{cochrane2022single}, who developed a principal-component analysis (PCA) framework for single- and multiple-flyby magnetometric detection of a Triton ocean, and showed that ionospheric contributions, external‑field variability, and trajectory noise can substantially degrade ocean detectability in single‑ and multiple‑flyby geometries.}

One of the pivotal goals of space magnetic field measurements is to decode the internal structure of Triton. However, factors influencing the distribution of the induced magnetic field include not only the subsurface ocean within Triton but also its ionospheric space environment and complex interactions with Neptune's magnetosphere. Therefore, in the next section, we will focus on the impact of interactions between Triton and Neptune's magnetosphere on the structure of space magnetic fields.

\subsection{\rev{Iron-core sensitivity}}

\rev{We evaluate the effects of an iron core and ocean composition using one iron-core model and two iron-core-free models. Table~\ref{tab:core_induction} reports the equatorial surface amplitude of the equivalent degree-one induced dipole, \(B_{\rm eq}^{\rm ind}(R_T)\), and the direction of the induced dipole axis in the Triton-centered TPhiO frame. Compared with the iron-core-free MgSO$_4$ reference model, the iron-core model changes \(B_{\rm eq}^{\rm ind}(R_T)\) from 2.58 to 2.48~nT, \(\theta\) from \(73.7^\circ\) to \(73.4^\circ\), and \(\phi\) from \(22.3^\circ\) to \(20.6^\circ\). Thus, the iron core produces only a small change in the degree-one induction response for the cases examined here. The iron-core-free Seawater endmember gives \(B_{\rm eq}^{\rm ind}(R_T)=2.29\)~nT and \(\phi=9.0^\circ\), indicating that ocean composition has a larger effect than the iron core in these sensitivity cases.}

%% 【HCP 2026-08-23】
\begin{table}[htbp]
\centering
\caption{\rev{Equivalent degree-one induced-dipole equatorial surface
amplitude and orientation for Triton models with and without an iron core.}}

\label{tab:core_induction}
\small
\setlength{\tabcolsep}{5pt}
\renewcommand{\arraystretch}{1.2}

\begin{tabularx}{\textwidth}{
    >{\raggedright\arraybackslash}p{0.20\textwidth}
    >{\raggedright\arraybackslash}X
    >{\centering\arraybackslash}p{0.18\textwidth}
    >{\centering\arraybackslash}p{0.08\textwidth}
    >{\centering\arraybackslash}p{0.08\textwidth}
}
\toprule
Interior model &
Ocean composition &
%%【注释，jhsept, 20260828】Surface induced field (nT) &
Equivalent dipole equatorial amplitude (nT) &
\multicolumn{2}{c}{Dipole orientation (TPhiO frame)} \\
\cmidrule(lr){4-5}
& & & $\theta$ ($^\circ$) & $\phi$ ($^\circ$) \\
\midrule

Iron core & MgSO$_4$, 10~g~kg$^{-1}$ & 2.48 & 73.4 & 20.6 \\

No iron core & MgSO$_4$, 10~g~kg$^{-1}$ & 2.58 & 73.7 & 22.3 \\

No iron core & Seawater, 10~g~kg$^{-1}$ & 2.29 & 73.4 & 9.0 \\

\bottomrule
\end{tabularx}
\end{table}

\section{MHD Simulation of Interaction Between Triton and the Magnetosphere of Neptune}

\subsection{Model Setup}

%% 【注释】In the aforementioned model, the induced magnetic field arises from the interaction between Triton's subsurface high-conductivity layer and its ionosphere under the influence of an external magnetic field.
%% 【注释】However, this model neglects the interaction between Neptune's magnetospheric plasma and Triton, and thus fails to account for the magnetic field contribution from the space current system. This simplification would lead to significant discrepancies between the theoretically calculated induced magnetic field and the detector-measured results. Observations and magnetohydrodynamic simulations of Europa have confirmed the existence of this issue\citep{harris2021multi}. Therefore, to more accurately simulate the space magnetic field around Triton, it is necessary to introduce an MHD simulation of the interaction between Triton and Neptune's magnetosphere.

In the model described above, the magnetic response is determined by electromagnetic induction associated with Triton’s conductive subsurface ocean and ionosphere under a time-varying external magnetic field. However, this approach does not account for the interaction between Neptune’s magnetospheric plasma and Triton, and therefore neglects magnetic perturbations generated by the surrounding plasma-current system. Such contributions may lead to differences between the idealized induction response and the magnetic field measured by a spacecraft. Similar effects have been demonstrated by observations and magnetohydrodynamic simulations of Europa \citep{harris2021multi}. To obtain a more complete description of the magnetic environment around Triton, we therefore include an MHD simulation of its interaction with Neptune’s magnetosphere.

%%【注释】 This MHD simulation relies on key physical parameters provided by the aforementioned \rev{PlanetProfile} and MoonMag models: the conductivity distribution of Triton and the equivalent magnetic moment of the induced magnetic field. The \rev{PlanetProfile} model provides the internal conductivity distribution of Triton. \rev{MoonMag uses this profile and the selected external period to calculate the equivalent induced magnetic moment. The steady SWMF run then applies the selected equivalent moment as an internal dipole field while calculating the surrounding plasma interaction.} Considering that the boundary condition of the MHD simulation is a steady field, the induced magnetic moment will be introduced as Triton's intrinsic magnetic moment in the simulation, thereby indirectly incorporating the time-varying effect of the external magnetic field and constructing a more self-consistent space electromagnetic environment model.

The MHD simulation is coupled to the MoonMag calculations through Triton’s equivalent induced magnetic moment. Because the MHD simulation assumes steady boundary conditions, this equivalent induced moment is implemented as an internal dipole field in each steady-state simulation. In this way, the induction response derived from the time-varying external field is incorporated into the plasma-interaction model, allowing the magnetic perturbations associated with both subsurface induction and magnetospheric plasma currents to be evaluated within a common framework.

%% 【注释】In terms of MHD modeling, we use the multi-fluid MHD simulation component under the Space Weather Modeling Framework (SWMF), which includes three ion species: nitrogen ions ($N^{+}$) in Neptune's magnetosphere, nitrogen molecular ions ($N_{2}^{+}$) and nitrogen ions ($N^{+}$) produced by the dissociation of Triton's neutral atmosphere ($N_{2}$). The density, momentum, and pressure of each ion species satisfy the continuity equation, momentum equation, and energy (or pressure) equation, respectively. In addition, the model considers the interactions between neutral gas ($N_{2}$), electrons, and the three ion species, introducing the corresponding source and loss terms into the MHD equations\citep{rubin2015self}. Therefore, we need to consider the governing equations for the mass, momentum, and energy (or pressure) of ions, the governing equation for the energy (or pressure) of electrons, and the governing equations for the evolution of the magnetic and electric fields.

For the plasma interaction, we use the multi-fluid MHD capability of SWMF. The simulation includes three ion fluids: magnetospheric N$^{+}$ and atmospheric N$_2^{+}$ and N$^{+}$ produced from Triton’s neutral N$_2$ atmosphere. Each ion species is described by its own continuity, momentum, and pressure equations. The model also includes interactions among neutral N$_2$, electrons, and the ion populations through appropriate source and loss terms \citep{rubin2015self}. The full system therefore consists of the governing equations for ion mass, momentum, and pressure, the electron pressure equation, and the evolution equations for the electromagnetic fields.

\subsection{Governing Equations of the Model}

The plasma interaction is described using the self-consistent multifluid MHD formulation implemented in SWMF \citep{rubin2015self}. For each ion species $s$, the mass, momentum, and pressure equations are

\begin{equation}
\frac{\partial \rho_s}{\partial t} + \nabla \cdot (\rho_s \mathbf{u}_s) = \frac{\delta \rho_s}{\delta t},
\label{eq:mass_equation}
\end{equation}

\begin{equation}
\frac{\partial (\rho_s \mathbf{u}_s)}{\partial t} + \nabla \cdot (\rho_s \mathbf{u}_s \mathbf{u}_s + I p_s) - Z_s e \frac{\rho_s}{m_s} (\mathbf{E} + \mathbf{u}_s \times \mathbf{B}) - \rho_s \mathbf{g} = \frac{\delta (\rho_s \mathbf{u}_s)}{\delta t},
\label{eq:momentum_equation}
\end{equation}

\begin{equation}
\frac{\partial p_s}{\partial t} + (\mathbf{u}_s \cdot \nabla) p_s + \gamma p_s (\nabla \cdot \mathbf{u}_s) = \frac{\delta p_s}{\delta t}, 
\label{eq: ion_pressure_equation}
\end{equation}
where $\rho_s$, $\mathbf{u}_s$, $p_s$, $m_s$, and $Z_s$ denote the mass density, bulk velocity, pressure, particle mass, and charge state of species $s$, respectively, and $\gamma=5/3$. The source and loss terms on the right-hand sides account for ionization, recombination, charge exchange, and collisional transfer of mass, momentum, and energy between plasma and neutral species.

\rev{In this multi-fluid MHD formulation, each ion species retains its own
momentum equation, including the electric and Lorentz forces. Electron inertia
is not solved as a separate momentum equation at the modeled MHD scales, but
the electron pressure is evolved separately because electron-impact
ionization, recombination, collisional energy exchange, and field-aligned heat
transport directly affect the electron energy budget:}

\begin{equation}
\frac{\partial p_e}{\partial t} + (\mathbf{u}_e \cdot \nabla) p_e + \gamma p_e (\nabla \cdot \mathbf{u}_e) + (\gamma - 1) \nabla \cdot (\frac{\boldsymbol{h}_e \cdot \boldsymbol{B}}{|\boldsymbol{B}|^2} \boldsymbol{B}) = \frac{\delta p_e}{\delta t},
\label{eq: electron_pressure_equation}
\end{equation}
where $\mathbf{h}_e$ denotes the electron heat flux. The source term includes electron energy gains and losses associated with ionization, recombination, and collisional energy exchange.

The magnetic field evolves according to Faraday's law,
\begin{equation}
\frac{\partial \mathbf{B}}{\partial t} = -\nabla \times \mathbf{E},
\label{eq: Faraday_law_equation}
\end{equation}
while the electric field is obtained from the generalized Ohm's law,
\begin{equation}
\mathbf{E} = -\mathbf{u}_e \times \mathbf{B} - \frac{1}{n_e e} \nabla p_e + \eta \mathbf{j},
\label{eq: Ohm_law_equation}
\end{equation}
where $n_e$ is the electron number density, $\eta$ is the resistivity, and $\mathbf{j}$ is the current density. The three terms on the right-hand side represent the motional electric field, the electron pressure-gradient contribution, and resistive effects, respectively. Together, these equations describe the coupled evolution of the ion fluids, electrons, and electromagnetic fields in Triton's plasma environment.

%%【待续，2025-07-16,03:57 pm】
\subsection{Boundary Conditions of the Model}

The outer boundary conditions of the simulation domain are defined as follows. Magnetospheric nitrogen ions are injected at the upstream boundary of the simulation domain, and their parameters are set according to the measurement results of the Voyager 2 probe, such as the number density, bulk velocity, and temperature of the magnetospheric plasma. These boundary conditions provide the initial plasma and magnetic field environment for the simulation. The downstream boundary of the simulation domain is of the outflow type, allowing plasma and magnetic fields to flow out of the simulation domain without reflection, simulating the extension of the plasma in the wake region of the icy moon. This setting avoids the interference of artificial reflected waves at the boundary on the internal flow field. Boundaries in other directions are set to ``float'', allowing plasma and magnetic fields to adjust freely without fixed constraints, thereby simulating Triton’s local interaction within the ice giant’s large-scale magnetospheric environment.

We consider the Triton's surface boundary conditions as follows. For plasma flowing toward the surface, inflow is allowed; when the flow direction is away from the surface, its velocity is set to 0, requiring no plasma outflow. The number densities of different ion species on the surface are fixed at initial values. For the magnetic field, a floating boundary is adopted, i.e., the gradient of the magnetic field perturbation (the difference between the total field and the induced dipole field) at Triton's surface is set to zero. This setting of the magnetic field boundary conditions approximately ignores the diffusion process of the magnetic field into the interior of the icy moon (such as the ice shell and ocean). In the future, it will be necessary to improve the magnetic field boundary conditions to accurately simulate the diffusion process of the magnetic field within the icy moon.

\subsection{Initial Conditions of the Model}

Initial conditions for plasma parameters are specified as follows. The number density of $N^{+}$ in Neptune's magnetosphere is set to $1.5\times10^{-3}$ $cm^{-3}$. It is assumed that the number densities of $N^{+}_{2}$ and $N^{+}$ (Triton-derived) contributed by Triton are one ten-thousandth of that of $N^{+}$ in Neptune's magnetosphere ($1.5\times 10^{-7}$ $cm^{-3}$), and the ion temperatures are all 100 eV\citep{hansen2021triton}. In future simulations, it will be necessary to optimize the settings based on the results of the Triton orbital mission. According to the observations of Voyager 2, the relative velocity between Triton and the ions in Neptune's magnetosphere is 43 km\,s$^{-1}$\citep{hansen2021triton}. Therefore, the upstream plasma velocity is set to 43 km\,s$^{-1}$. It is assumed that the electron thermal pressure is consistent with the total thermal pressure of the three ion species.

The initial settings of the magnetic field include the background magnetic field in Neptune's magnetosphere and the approximate dipole-induced magnetic field inside Triton. In the TPhiO reference frame centered on Triton, the background magnetic field is $B_x=1.45\times 10^{-4}$~nT, $B_y=0.206$~nT, and $B_z=3.06$~nT. For the 10~g~kg$^{-1}$ MgSO$_4$ ocean model, MoonMag predicts an induced dipole magnetic field of 2.44~nT at the magnetic equator, with the magnetic axis oriented at approximately $\theta=73^\circ$ and $\phi=18^\circ$. These values differ slightly from those in Table~\ref{tab:core_induction} because this calculation uses fewer radial layers to represent Triton's interior structure. \rev{The calculation represents the interaction state associated with this selected induced dipole and upstream condition. A time-dependent flyby calculation would additionally follow the forcing phase along the trajectory.}

%% 【注释】The neutral gas background is initially defined as follows. At an altitude of 575 km above Triton's surface, the temperature of $N_2$ is approximately 100 K, and the particle number density is approximately $4\times 10^8$ $cm^{-3}$\citep{krasnopolsky1993temperature}. The atmospheric pressure on Triton's surface is approximately 1 Pa. By comparing with the corresponding parameters of the Earth, in our MHD simulation, the number density of $N_2$ on Triton's surface is set to $10^{15}$ $cm^{-3}$, with a scale height of 35 km. The neutral gas background participates in the simulation of interactions between plasma and neutral gas as an initial condition.

The neutral atmosphere is initialized as follows. At an altitude of 575 km above Triton’s surface, the N$_2$ temperature is approximately 100 K and the number density is about $4\times10^{8}$ cm$^{-3}$ \citep{krasnopolsky1993temperature}. Given Triton’s surface atmospheric pressure of approximately 1 Pa, we adopt an N$_2$ number density of $10^{15}$ cm$^{-3}$ at the surface and a scale height of 35 km in the MHD model. This prescribed neutral background serves as the initial atmospheric state and participates in plasma-neutral interactions throughout the simulation.

\subsection{Simulation Results}

\begin{figure}[htbp]%
\centering
\includegraphics[width=1.0\textwidth]{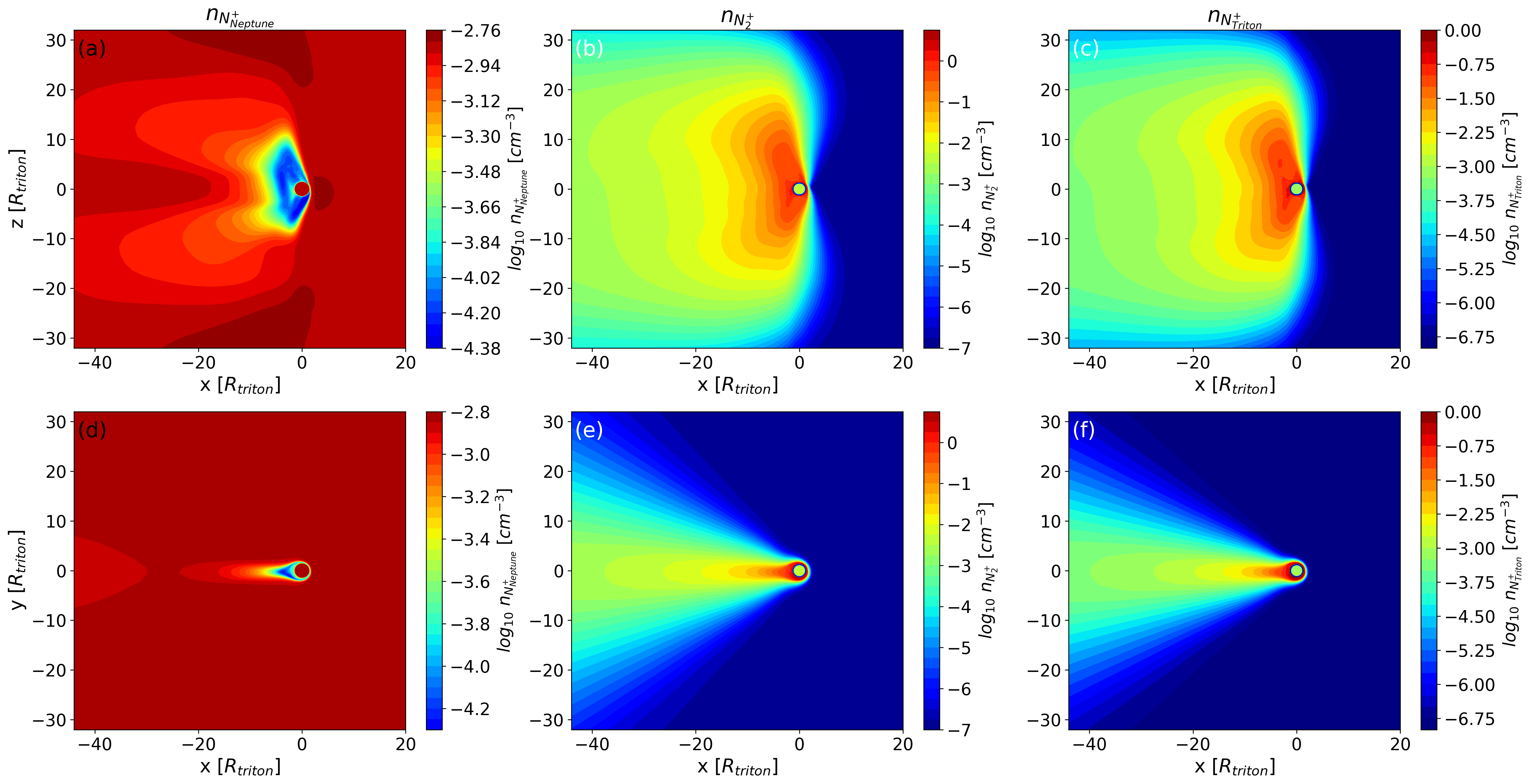}
\caption{Number density distributions of three ion species—nitrogen ions ($N^{+}$) in Neptune's magnetosphere, Triton-derived nitrogen molecular ions ($N_{2}^{+}$), and Triton-derived nitrogen ions ($N^{+}$)—in the x-z and x-y planes.
}\label{fig5}
\end{figure}

\begin{figure}[htbp]%
\centering
\includegraphics[width=1.0\textwidth]{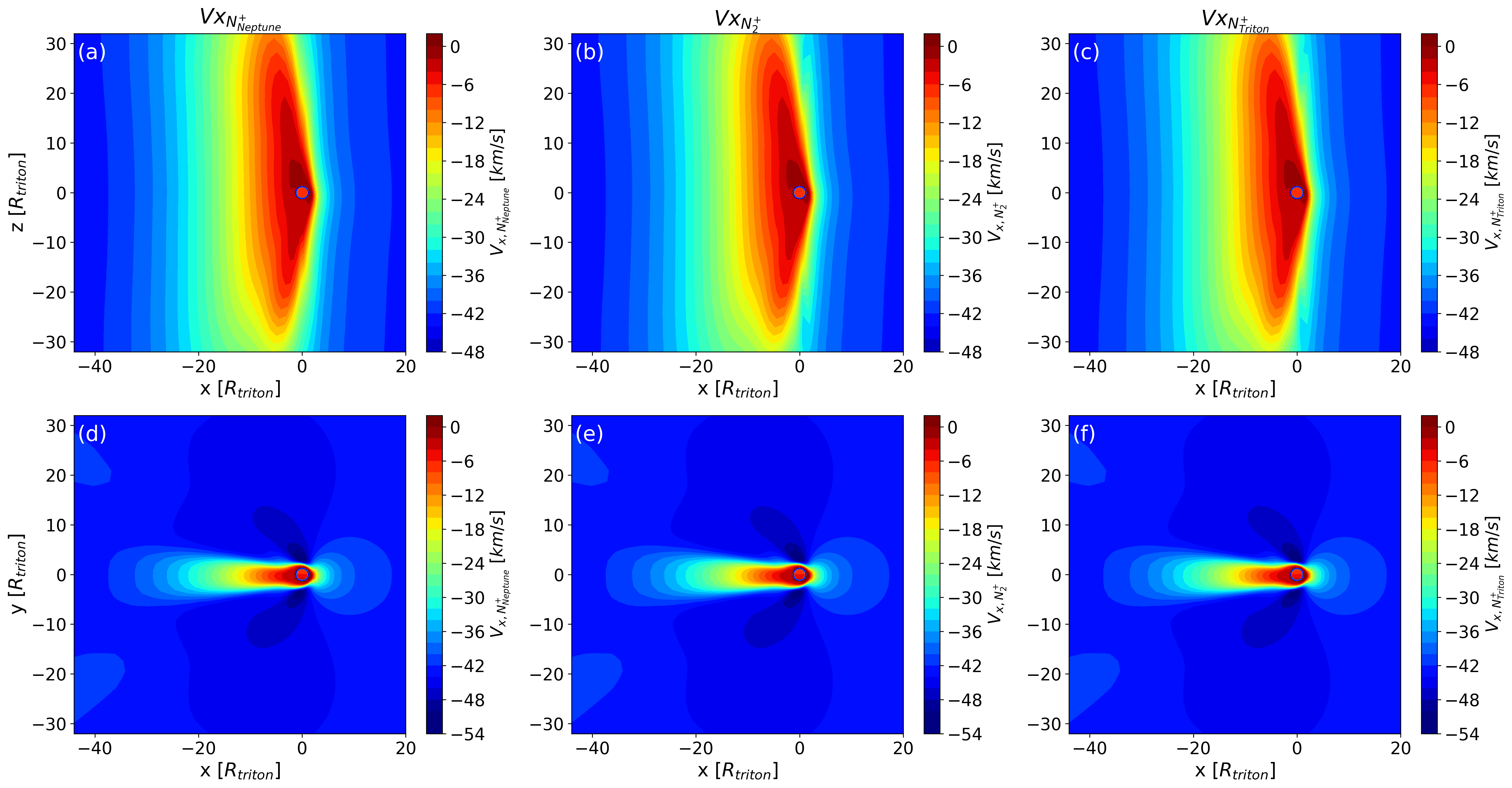}
\caption{Velocity distributions of three ion species—nitrogen ions ($N^{+}$) in Neptune's magnetosphere, Triton-derived nitrogen molecular ions ($N_{2}^{+}$), and Triton-derived nitrogen ions ($N^{+}$)—in the x-z and x-y planes.}\label{fig6}
\end{figure}

Figures~\ref{fig5} and \ref{fig6} show the number density and bulk velocity distributions of three ion species: nitrogen ions ($N^{+}$) in the Neptune's magnetosphere, Triton's nitrogen molecular ions ($N^{+}_{2}$), and Triton's nitrogen ions ($N^{+}$) in the $X$-$Z$ and $X$-$Y$ planes. It can be seen from Figure~\ref{fig5} that, due to the interaction between $N^{+}$ ions and Triton's neutral atmosphere, an obvious low-density cavity region and a velocity reduction region are formed around Triton. At the same time, photoionization and impact ionization processes ionize neutral nitrogen molecules ($N_{2}$) to generate $N^{+}_{2}$, which is distributed around Triton and extends downstream along the background flow. In the $X$-$Z$ plane, since the background magnetic field is mainly distributed along the $Z$-direction, the interaction between the background ion flow and Triton forms a typical Alfv\'en wing structure, showing a nearly axisymmetric distribution (Figure~\ref{fig6}). Because the background flow velocity is much lower than the Alfv\'en velocity ($\approx443\ \mathrm{km\,s^{-1}}$, calculated from background magnetic field $\approx 3~\mathrm{nT}$ and number density of $N^{+}$ $\approx 10^{-2.8}\mathrm{cm}^{-3}$), the angle between the Alfv\'en wing and the $Z$-axis is very small.

\begin{figure}[htbp]%
\centering
\includegraphics[width=1.0\textwidth]{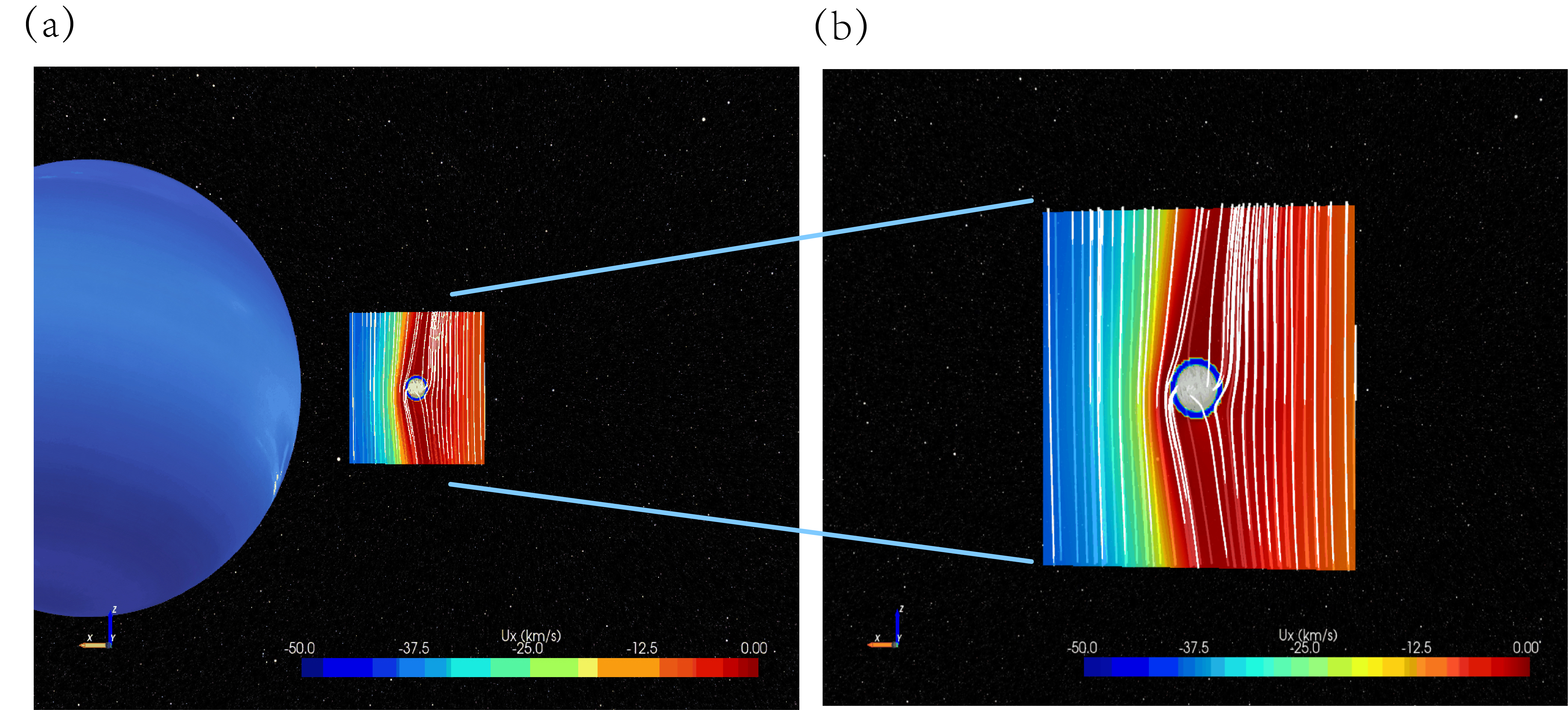}
\caption{\rev{Velocity distribution and magnetic-field lines for magnetospheric nitrogen ions ($N^{+}$). (a) Global context of the Triton interaction region in Neptune's system. (b) Enlarged view of the flow and field-line structure around Triton. The color scale gives the $x$-component of velocity and the white curves show magnetic-field lines. (See Supplementary Animation 1.)}}\label{fig7}
\end{figure}

Furthermore, we superimpose the magnetic field lines on the bulk velocity distribution in the $X$-$Z$ plane, and it can be seen that the direction of the magnetic field lines is basically consistent with the plasma bulk velocity distribution (Figure~\ref{fig7}). \rev{The global panel locates the simulated interaction region relative to Neptune, and the enlarged panel resolves the flow deflection and field-line structure close to Triton.} Some magnetic field lines connect to the Triton's surface, and these connection positions on the Triton's surface correspond to the positive and negative pole regions of Triton's induced magnetic field.

%% 【注释】Animation 1. Velocity distribution of nitrogen ions (N⁺) in Neptune's magnetosphere in the x-z plane, with white solid lines representing magnetic field lines.

\begin{figure}[htbp]%
\centering
\includegraphics[width=1.0\textwidth]{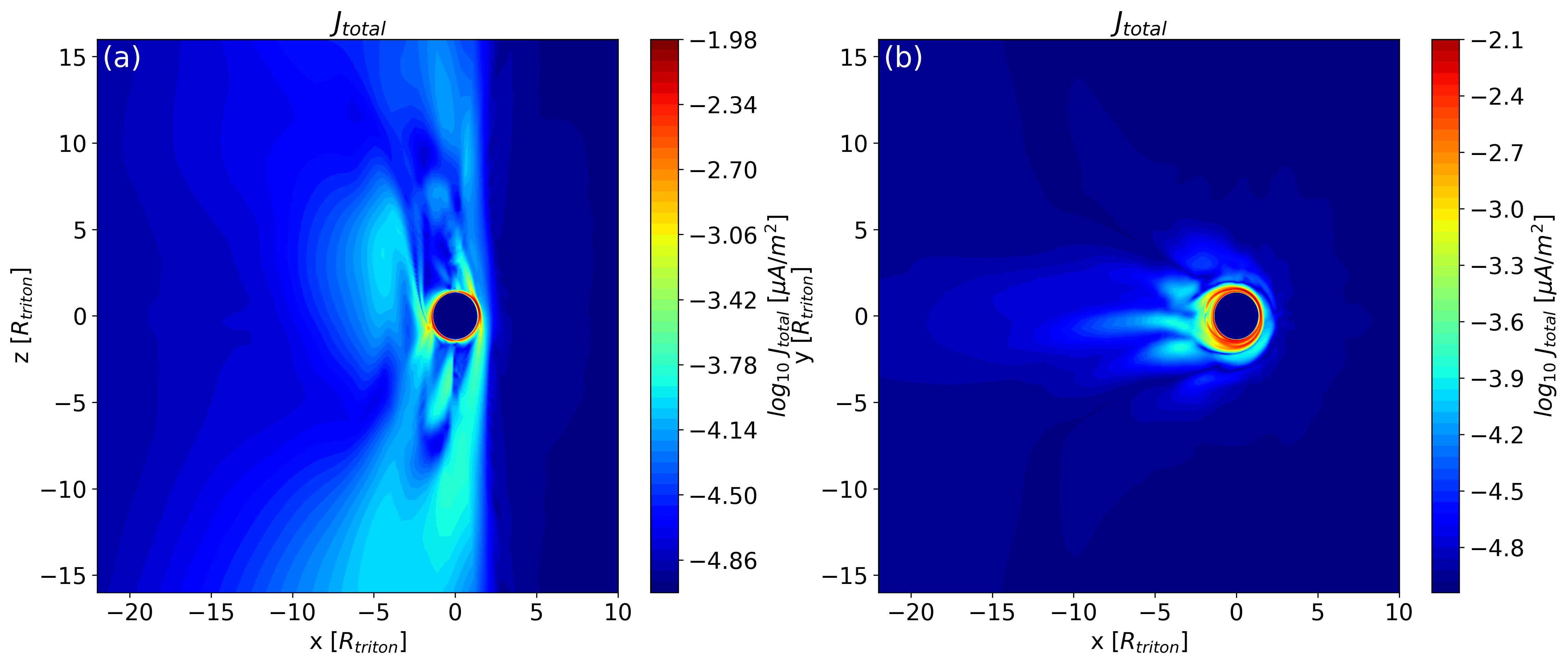}
\caption{(a) Spatial distribution of total current density in the x-z plane; (b) Spatial distribution of total current density in the x-y plane.}\label{fig8}
\end{figure}

Figure~\ref{fig8} shows the distribution of space electric currents around Triton obtained through MHD simulation. The space magnetic field near Triton consists of three parts: the background magnetic field provided by Neptune's magnetosphere, the induced magnetic field generated by Triton's internal high-conductivity ocean, and the magnetic field generated by the space electric currents around Triton. To calculate the magnetic field distribution contributed by space electric currents, we solve the following equation. Under the Coulomb gauge ($\nabla\cdot \mathbf{A}=0$), Ampère's law in Maxwell's equations can be simplified as:
\begin{equation}
\nabla^2 \mathbf{A} = -\mu_0 \mathbf{J}
\label{eq: Ampere_law_for_A}
\end{equation}
where $\mathbf{A}$ is the magnetic vector potential, $\mathbf{J}$ is the current density, and $\mu_{0}$ is the vacuum permeability. This equation is in the form of a typical Poisson equation and can be solved numerically using the finite difference method (FDM).

After acquiring the spatial distribution of the current density $\mathbf{J}$, solving for the magnetic vector potential $\mathbf{A}$ allows further calculation of the magnetic field, $\mathbf{B}=\nabla\times\mathbf{A}$.
Through this process, we can calculate the magnetic field generated from the simulated spatial current distribution. Future studies should examine the sensitivity of the inferred magnetic field to the current distribution, plasma parameters, and numerical assumptions.

\begin{figure}[htbp]%
\centering
\includegraphics[width=1.0\textwidth]{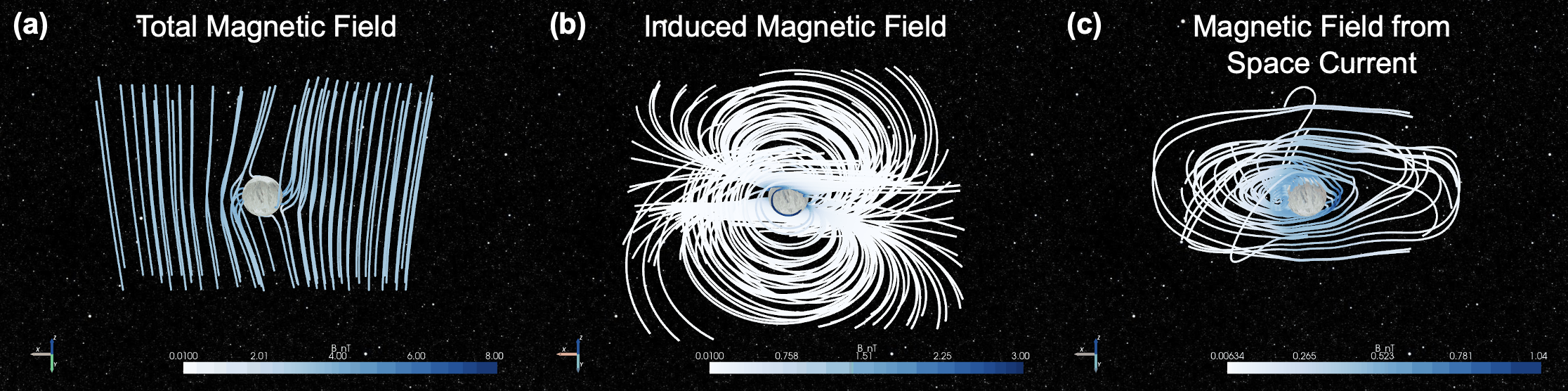}
\caption{(a) Spatial distribution of magnetic field lines from MHD simulations; (b) Spatial distribution of the induced magnetic field calculated using the MoonMag model; (c) Spatial distribution of the magnetic field corresponding to space electric currents from MHD simulations. (For time variations of the background magnetic field and induced magnetic field, see Supplementary Animation 2.)}\label{fig9}
\end{figure}

The results of the MHD simulation show that the total magnetic field is mainly distributed along the $Z$-direction (Figure~\ref{fig9}a), similar to the initial configuration ($B_x=1.45\times 10^{-4}$~nT, $B_y=0.206$~nT, and $B_z=3.06$~nT). The induced magnetic field generated by the internal conductive ocean presents an approximate dipole characteristic (Figure~\ref{fig9}b), whereas the magnetic field caused by the space electric currents around Triton has a more complex distribution (Figure~\ref{fig9}c). A further comparison reveals that the intensity of the magnetic field generated by space electric currents is of a similar order of magnitude to the induced magnetic field caused by the subsurface ocean. This suggests that space electric currents have a significant impact on the magnetic field structure surrounding Triton. Therefore, measuring space electric currents is indispensable, and their accuracy will directly affect the reliability of the inversion results for Triton's internal structure.

\rev{The MHD result depends on the selected upstream plasma, neutral atmosphere, chemical processes, boundary conditions, and applied induced dipole. Simultaneous measurements of plasma density, composition, velocity, and temperature would therefore provide the information needed to model the current system during a future magnetic sounding experiment.}

%% 【注释】Animation 2. Time variations of Neptune's magnetospheric magnetic field and Triton's induced magnetic field as observed at Triton.

%%【待续，2025-07-19, 11:30 am】

\begingroup
\color{revisionblue}

\section{Seismic Forward Modeling and Inversion of Triton's Ice Shell and Ocean}
\label{sec:seismic_inversion}

\subsection{Purpose of the seismic experiment}
\label{subsec:seismic_constraints}

The magnetic response calculated in Section~3 depends jointly on ocean thickness, depth, and electrical conductivity. Similar magnetic amplitudes and phases can therefore be produced by different thickness--conductivity combinations. Seismic travel times provide a complementary observable because reflections and transmissions respond directly to the depths of the ice--ocean and ocean--rock interfaces. We use a controlled numerical experiment to determine which parts of Triton's layered structure can be recovered from such arrivals.

The experiment contains two linked tests. First, a three-dimensional forward calculation generates synthetic seismograms for a known Triton model. The synthetic records are then treated as \rev{virtual measurement data}, and a travel-time inversion attempts to recover the source and layer parameters. The inversion does not prescribe the true horizontal source location or layer thicknesses, while retaining explicit priors on shallow source depth and ocean P-wave velocity. This recovery test evaluates whether the selected phases contain independent information on ice and ocean thickness. Second, in Section~\ref{sec:seismic_detectability}, we compare the amplitudes of the synthetic signals with a low-frequency band noise benchmark. This additional step is required because a numerically visible arrival can contribute to an inversion only when its amplitude exceeds the noise level of a deployed instrument.

The simulation quantifies the main travel-time sensitivities under controlled conditions. It uses a spherically symmetric interior, noise-free phase picks, and a small five-station network. %% 【注释】Attenuation, small-scale scattering, topography, rotation, ellipticity, self-gravitation, lander coupling, and environmental noise are reserved for later ensemble calculations. 
The recovered errors therefore describe this specific end-to-end synthetic test. We refer to the combined three-dimensional spectral-element forward modeling, polarization-assisted phase identification, and hierarchical travel-time inversion workflow as `TritonSeis'.

\subsection{Interior model, source, and numerical configuration}
\label{subsec:seismic_model}

We calculate the wavefield with the regional configuration of SPECFEM3D\_GLOBE, which solves the coupled elastic--acoustic equations using the spectral-element method. The computational domain covers $30^{\circ}\times30^{\circ}$ and is centered at $0^{\circ}$ latitude and $120^{\circ}$ longitude. It employs a 256 × 256 horizontal spectral-element mesh distributed across 256 Message Passing Interface (MPI) processes in a 16 × 16 domain decomposition. A traction-free condition is applied at Triton's surface, and absorbing conditions are used at the lateral boundaries to reduce artificial reflections from the edge of the regional mesh.

The radial model is derived from the same PlanetProfile structure used for the electromagnetic calculation. Triton's radius is 1352.6~km. In the discretized seismic mesh, the ice--ocean interface is 113.75~km below the surface and the ocean--rock interface is at 238.55~km depth, giving an ocean thickness of 124.80~km. P- and S-wave velocities (\(V_P\) and \(V_S\)) and density vary radially in the solid layers. The liquid ocean has $V_S=0$, while $V_P$ increases from approximately 1.53~km\,s$^{-1}$ near the top to 1.75~km\,s$^{-1}$ near the seafloor.

A controlled moment-tensor source is placed at $0^{\circ}$ latitude, $120^{\circ}$ longitude, and 5~km depth. Its scalar moment is $8.66\times10^{16}$~N\,m, corresponding to $M_w=5.23$, and its half-duration is 1.6~s. The diagonal-dominated tensor has a nonzero trace and radiates predominantly compressional energy. This source provides clear P-wave illumination of the two interfaces and allows the recovery method to be tested with a well-defined wavefield.

Five three-component stations are placed at epicentral distances of approximately 97-255~km (Figure~\ref{fig10}a). Each station records 9~min of radial ($R$), transverse ($T$), and vertical ($Z$) velocity. The records are band-pass filtered between 0.05 and 0.25 Hz before phase picking. %% 【注释】This band was selected as a compromise between instrument sensitivity, structural scale, and propagation robustness. The 0.05--0.25~Hz range lies within the long-period operating band of VBB planetary seismometers such as InSight SEIS \citep{mimoun2017noise,lognonne2019seis} and is also used consistently for the noise estimate in Section~5.5. At 0.25~Hz, the corresponding P-wave wavelengths are approximately 17~km in the ice shell and 6.4~km in the ocean, both smaller than the respective layer thicknesses. The low-frequency band also avoids relying on higher-frequency energy that would be more susceptible to attenuation, scattering, and small-scale ice heterogeneity in an icy shell \citep{StahlerEtAl2018}. Tests with nearby passbands produced consistent phase identifications and inversion results, indicating that the conclusions are not sensitive to modest variations in the filter corners. 
\rev{The upper limit of the frequency band is set by the reliable frequency range of the spectral-element mesh, while the lower limit ensures that the direct and reflected P-wave arrivals remain temporally separable.}
%%【注释，删去】SPECFEM3D\_GLOBE starts the output 2.4~s before the center of the source-time function; this known offset is included when the source-centroid time is estimated.

\begin{figure}[t]
\centering
\includegraphics[width=1.0\textwidth]{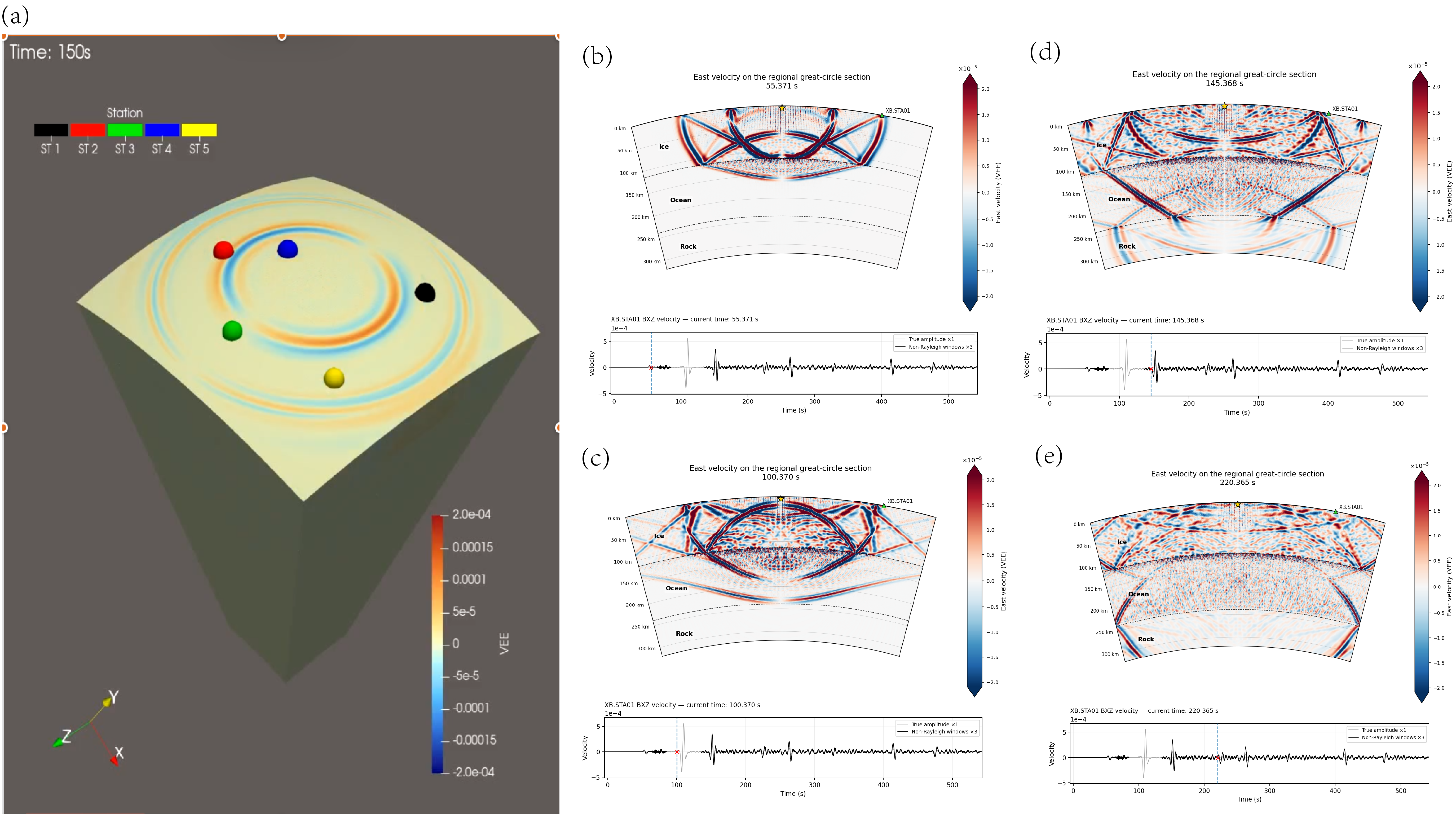}
\caption{Numerical configuration and seismic-wave propagation in the Triton reference model. (a) Regional computational domain showing the source and five surface stations. (b) Radial distributions of $V_P$, $V_S$, and density across the ice shell, ocean, and rocky interior. (c--e) Representative snapshots showing propagation through the ice, transmission into the ocean, and interaction with the ocean--rock interface.}
\label{fig10}
\end{figure}

\subsection{Wavefield characteristics and phase identification}
\label{subsec:phase_identification}

The first coherent arrival at each station is the direct P wave, which propagates mainly through the ice shell. Its moveout with distance is controlled primarily by the horizontal source location, source-centroid time, and effective ice-shell P-wave velocity. Later wave packets include a reflection from the ice--ocean boundary and energy that enters the ocean, interacts with the ocean--rock boundary, and returns to the surface. We denote the principal ice-bottom reflection by $P_iP$ and the later ocean-sensitive arrival by $P_{oc}P$.

The large decrease in $V_P$ and shear rigidity at the ice--ocean boundary generates a strong reflected P-wave field in the ice. Energy transmitted into the ocean with slower propagation velocity is refracted toward the vertical, travels through the water layer, and interacts with the rocky interior. Figure~\ref{fig11} compares the filtered vertical-component seismograms with theoretical travel-time curves calculated from the reference radial model. The ordering and moveout of the direct P, $P_iP$, and $P_{oc}P$ families follow their expected paths.

Arrival time alone can produce ambiguous assignments because weak reflections overlap with coda energy. Candidate peaks are therefore screened with three-component polarization. A P-wave candidate should be strongest on the vertical or radial component and should have particle motion approximately within the source--receiver great-circle plane. Peaks dominated by transverse motion or incoherent coda are removed. The retained candidates are then required to yield mutually consistent interface depths across the five stations. This combination of predicted moveout, polarization, and multi-station consistency reduces phase-association errors.

\begin{figure}[t]
\centering
\includegraphics[width=1.0\textwidth]{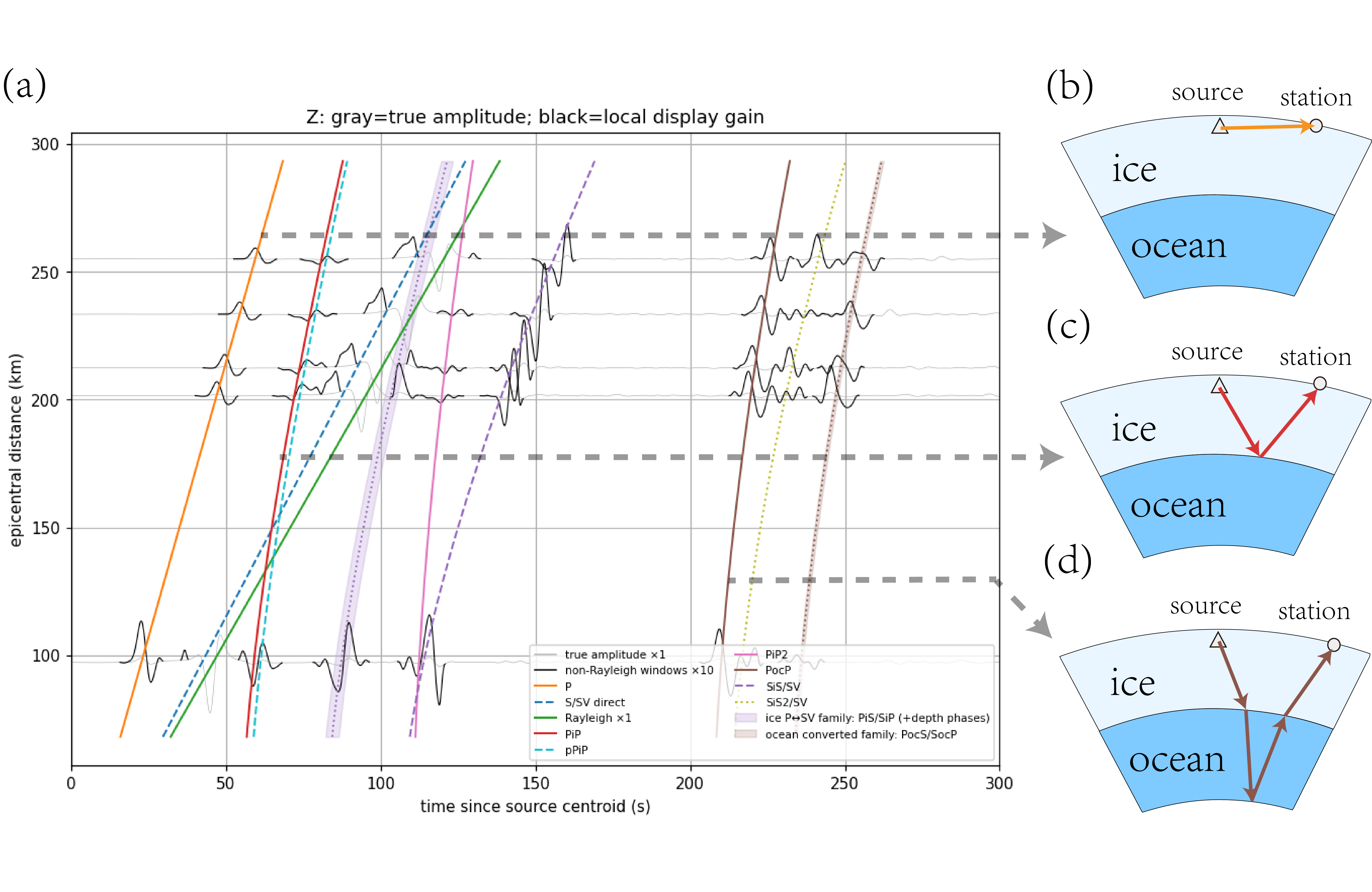}
\caption{Identification of P-wave phases in the synthetic Triton seismograms. (a) Filtered vertical-component records as a function of epicentral distance. Gray traces show the original amplitudes, black traces display locally amplified weaker arrivals, and colored curves show theoretical travel times. (b--d) Schematic paths of the direct P wave, the $P_iP$ reflection from the ice--ocean interface, and the ocean-sensitive $P_{oc}P$ phase interacting with the ocean--rock interface. All records are filtered between 0.05 and 0.25~Hz.}
\label{fig11}
\end{figure}

\subsection{Hierarchical travel-time inversion and synthetic recovery test}
\label{subsec:hierarchical_inversion}

For phase $k$ at station $i$, the predicted arrival time is written as
\begin{equation}
t_{i,k}^{\rm pred}=\Delta t_{\rm origin}+T_k(\boldsymbol{x}_s,z_s,\boldsymbol{x}_i,\boldsymbol{m}),
\label{eq:predicted_arrival}
\end{equation}
where $\boldsymbol{x}_s$ and $z_s$ are the source location and depth, $\boldsymbol{x}_i$ is the station location, and $\boldsymbol{m}$ contains the effective velocities and layer thicknesses. The parameters are estimated by minimizing the residuals between picked and predicted arrival times at all stations.

%%【注释】The inversion is divided into three stages so that source uncertainty is reduced before the deeper interfaces are estimated.
\rev{The inversion follows a three‐stage workflow, where source‐related uncertainties are first mitigated before constraints on the deeper subsurface interfaces are derived.}

%%【注释】\textbf{Stage 1: source and effective ice velocity.} 
\textbf{Stage 1 estimates the source location, the source-centroid time, and effective ice‑shell velocity.}
The five direct-P arrivals are used to estimate the horizontal source location, source-centroid time, and a single effective $V_P$ for the ice shell. 
%%【注释】Simultaneously solving for depth, velocity, and origin time is poorly conditioned with five stations, so a shallow-source constraint is applied and the inversion depth is fixed at the surface. The 5 km difference from the input source is retained as one contributor to the recovery error.
%%【注释，jshept, 20260826】Simultaneously solving for source depth, velocity, and origin time is poorly conditioned for the present five-station geometry, so $z_s$ is fixed at the surface. For a shallow source, direct-P travel times are only weakly sensitive to source depth. In a local homogeneous approximation, letting $D_i$ denote the epicentral distance to station $i$, $\left|\partial T_i/\partial z_s\right|/\left|\partial T_i/\partial D_i\right|=z_s/D_i$. For $z_s=5$~km and $D_i\simeq97$--255~km, this ratio is only 0.05--0.02, corresponding to a travel-time difference of approximately 0.03--0.01~s relative to a surface source \citep{AkiRichards2002,Bose2017}. A shallow-source prior is also physically plausible because ice-shell seismicity is expected preferentially in the cold, brittle upper shell, whereas warmer ice at depth deforms more ductilely \citep{PanningEtAl2018,VanceEtAl2018}, consistent with the stagnant-lid thermal structure obtained in Section~2.1. Deeper sources cannot be excluded; the zero-depth constraint is therefore treated as an effective shallow-source approximation, and the 5 km difference from the input source is retained as part of the recovery error.
\rev{Simultaneously solving for depth, velocity, and origin time is poorly conditioned with five stations, since arrival‐time data from this limited‐aperture network provides weak sensitivity to source depth. A shallow‐source constraint is therefore applied, and the inversion depth is fixed at the surface. This shallow-source assumption is physically plausible: ice-shell seismicity is expected preferentially in the cold, brittle upper shell, whereas warmer ice at depth deforms ductilely \citep{PanningEtAl2018,VanceEtAl2018}, consistent with the stagnant-lid thermal structure in Section~2.1. This known 5‐km offset from the input source depth acts as a systematic bias propagating into subsequent parameter estimates.}

%%【注释】 \textbf{Stage 2: ice-shell thickness.} With the Stage-1 parameters fixed, local maxima following the direct P arrival are tested as candidate $P_iP$ reflections. Each candidate time maps to a trial ice thickness. The preferred thickness is selected from the cluster that gives the smallest combined residual and the greatest consistency among all five stations.

%%【stage2 is modified by 谢浩恩， 2026.8.26 】
%% 【注释】\textbf{Stage 2: ice-shell thickness.}
\rev{\textbf{Stage 2 solves for ice‑shell thickness.} Only post-direct-$P$ peaks showing linear, $P$-wave-like polarization in the radial--vertical plane are retained as candidates. With the Stage-1 source parameters and effective ice $V_P$ fixed, each candidate is interpreted as $P_iP$ and mapped to a trial ice thickness using the $P_iP$ travel-time relation for a spherically symmetric ice shell. The defining criterion is cross-station consistency: a genuine $P_iP$ yields a common thickness across all five stations, whereas coda or unrelated phases do not. Trial thicknesses are therefore scored by both the number of supporting stations (within a tolerance window) and the amplitude-weighted tightness of the candidate cluster. The solution is taken from the thickness range with maximum station coverage; if multiple coherent clusters remain, the shallowest is selected as the primary $P_iP$ to exclude intra-ice-shell multiples. The final ice-shell thickness is the amplitude-weighted mean of the station estimates in the selected cluster.}

\textbf{Stage 3: ocean thickness.} Later candidates are tested as $P_{oc}P$ arrivals. The time separation between $P_iP$ and $P_{oc}P$ contains the two-way propagation contribution of the ocean. The present geometry provides limited independent sensitivity to ocean velocity, so $V_{P,\rm ocean}=1.60$~km\,s$^{-1}$ is adopted from the reference-model range and the ocean thickness is estimated. This sequential procedure keeps the origin of each constraint visible and also shows how early-stage velocity errors propagate into deeper-layer estimates.

The recovered horizontal source position is 0.49~km from the input epicenter. The recovered source-centroid time differs from the input value by approximately 0.4~s. The effective ice $V_P$ is 4.403~km\,s$^{-1}$, which is 7.9\% above the path average through the ice. The direct-P rays bend toward and preferentially sample the faster shallow ice, so the recovered effective velocity exceeds the depth-averaged \(V_P\) through the ice shell.

The recovered ice thickness is 123.32~km, compared with the 113.75 km input, and is therefore 8.4\% larger. Applying the relatively high shallow-ice velocity to the full reflection path moves the inferred interface deeper. The recovered ocean thickness is 109.20 km, or 12.5\% smaller than the input value of 124.80 km. This underestimate partly reflects the overestimated ice-shell thickness, as the \(P_{oc}P\) travel time is partitioned between ice and ocean segments. 
%%【注释】Table~\ref{tab:synthetic_recovery} summarizes the complete recovery test.
Table~\ref{tab:synthetic_recovery} compiles all key input and recovered parameters from this end‐to‐end synthetic recovery test, together with the corresponding absolute
and relative errors.

\begin{table*}[t]
\centering
\caption{Recovery of source and structural parameters from the synthetic seismograms.}
\label{tab:synthetic_recovery}
\small
\begin{tabularx}{\linewidth}{@{}>{\raggedright\arraybackslash}Xcccc@{}}
\toprule
Parameter & Input or reference & Recovered & Error & Relative error\\
\midrule
Source latitude & $0.000^{\circ}$ & $-0.001^{\circ}$ & $-0.001^{\circ}$ & --\\
Source longitude & $120.000^{\circ}$ & $120.021^{\circ}$ & $+0.021^{\circ}$ & --\\
Source centroid time & 2.400~s & 2.011~s & $-0.389$~s & --\\
Effective ice $V_P$ & 4.289~km\,s$^{-1}$ & 4.403~km\,s$^{-1}$ & $+0.114$~km\,s$^{-1}$ & $+2.67\%$\\
Ice-shell thickness & 113.75~km & 123.32~km & $+9.57$~km & $+8.4\%$\\
Ocean thickness & 124.80~km & 109.20~km & $-15.60$~km & $-12.5\%$\\
\bottomrule
\end{tabularx}
\end{table*}

\rev{The recovery test is end-to-end for a single synthetic interior, source, and station-network geometry. Robust uncertainty quantification would require ensemble tests encompassing variable source mechanisms and depths, laterally heterogeneous ice, attenuation and scattering, realistic noise, phase-picking uncertainty, and alternative station layouts. The present test nevertheless demonstrates that the reflected $P$-wave families contain sufficient independent timing information to resolve the two interfaces to first order.}

\subsection{Estimate of Seismic Detectability}
\label{sec:seismic_detectability}
We make a simple estimate of the seismic-event magnitude detectable by a future Triton seismometer. We adopt a Very Broad Band (VBB) acceleration-noise level of $2.5\times10^{-9}\ {\rm m\,s^{-2}\,Hz^{-1/2}}$ as a reference for a flight-proven planetary seismometer. This value is based on the performance and noise requirements of the InSight Seismic Experiment for Interior Structure (SEIS) instrument \citep{mimoun2017noise,lognonne2019seis}. The Farside Seismic Suite (FSS), an InSight-heritage VBB payload scheduled for a NASA CLPS lunar landing in the Schr\"odinger basin, will provide an additional near-term reference for landed seismology on a moon-body analog~\citep{aboobaker2024farside}.

%%【1.26e-6 --> 7.96e-7；2.52e-9 --> 1.59e-9,因为频段从0.02-0.10Hz-->0.05-0.25Hz，谢浩恩，2026.8.27】
Over the frequency range of $0.05$--$0.25$ Hz used in the present simulation, this value corresponds to an rms velocity noise of approximately 
$1.59\times10^{-9}\ {\rm m\,s^{-1}}$. 
We adopt a signal‐to‐noise ratio of 5 as our detection criterion. To account for poorly constrained noise sources including ground‐coupling effects, lander‐induced vibration, thermal perturbations, and surface‐environmental noise, we apply a conservative factor of 100 to the reference VBB instrument‐noise level to approximate the overall effective noise floor. Combining this total noise floor with the SNR requirement yields a conservative velocity detection threshold of approximately \(7.96\times10^{-7}\ \mathrm{m\,s^{-1}}\).

Having established the detection threshold, we next estimate peak ground velocity for events of different magnitude and distance. We adopt as reference the filtered peak velocity of $\sim 10^{-4}\ \mathrm{m\,s^{-1}}$ from our $M_w \simeq 5.23$ synthetic source at 150~km epicentral distance, and assume that peak amplitude scales with seismic moment, decays geometrically as $1/R$, and is attenuated with an effective $Q=100$ (a typical value for icy-shell attenuation in feasibility studies). The scaling relation for arbitrary magnitude-distance pairs is

\begin{equation}
v_{\rm pk}(M_w,R)
=
v_{\rm ref}
10^{1.5(M_w-M_{w,\rm ref})}
\frac{R_{\rm ref}}{R}
\exp\left[
-\frac{\pi f_{\rm eff}(R-R_{\rm ref})}
{Q V_P}
\right],
\label{eq:seismic_detectability}
\end{equation}
where %%【注释：where要和上面公式紧挨着，不要隔一行；隔一行的话，where在pdf里就会新成一段，段首缩进。】
$v_{\rm ref}=10^{-4}\ {\rm m\,s^{-1}}$,
$M_{w,\rm ref}=5.23$,
$R_{\rm ref}=150$ km,
$f_{\rm eff}=0.05$ Hz, and
$V_P=4\ {\rm km\,s^{-1}}$.

Based on this scaling equation, we solve for the minimum moment magnitude at each epicentral distance by setting the predicted peak velocity equal to the detection threshold ($v_{pk}=v_{\rm threshold}$). The minimum magnitudes for detecting the primary (largest-amplitude) arrival are approximately \(M_w=3.78\), \(4.12\), and \(4.55\) at 100, 300, and 1000~km, respectively. Interface-reflected phases used to constrain the ice--ocean and ocean--rock boundaries may have considerably lower amplitudes; for sensitivity cases with interface phases at 10\% and 1\% of the primary amplitude, the required magnitudes increase by $\sim$0.67 and $\sim$1.33, respectively. These estimates are summarized in Table~\ref{tab:seismic_detectability}.

\begin{table}[htbp]
\centering
\caption{Estimated minimum moment magnitude for detecting the largest arrival and weaker interface-reflected phases under the adopted conservative noise condition.}
\label{tab:seismic_detectability}
\begin{tabular}{cccc}
%%【注释】\hline
\toprule
Epicentral distance
& Minimum magnitude
& 10\% interface phase
& 1\% interface phase \\
\midrule
100 km  & 3.78 & 4.44 & 5.11 \\
300 km  & 4.12 & 4.78 & 5.45 \\
1000 km & 4.55 & 5.21 & 5.88 \\
\bottomrule
\end{tabular}
\end{table}
%%【update the result of the table，因为频段从0.02-0.10Hz-->0.05-0.25Hz，谢浩恩，2026.8.27】

These values provide a first-order estimate under the adopted source, propagation, attenuation, and noise conditions. The calculation leaves Triton's seismic-event rate, source-mechanism distribution, environmental noise, and instrument coupling unconstrained. \rev{Follow‐up analyses could refine these estimates given observational constraints on surface noise, frequency‐dependent attenuation, and a broader range of plausible seismic‐source behaviors.}

\subsection{Single-station extension and joint interpretation}
\label{subsec:single_station}

A network improves source localization and phase association. A single station can still obtain structural information when source geometry is independently constrained or when differential times are used. For a located impact or repeated icequake, the time difference between a direct wave and an interface reflection cancels the unknown origin time. Similar single-station seismic approaches have been examined for Europa using natural seismic and acoustic sources~\cite{LeeEtAl2003}. If a fresh impact could be identified in surface‐imaging data, this impact would provide direct spatial constraints on source location. In the absence of such imaging constraints, waveform‐polarization analysis can only determine the back‐azimuth toward the seismic source.

Continuous ambient noise provides a second single-station approach. The autocorrelation of a long-duration seismic record,
\begin{equation}
C(\tau)=\int u(t)u(t+\tau)\,\mathrm{d}t,
\label{eq:noise_autocorrelation}
\end{equation}
can approximate the zero-offset near-vertical reflection response beneath the receiver when the wavefield contains a sufficiently broad angular distribution of propagating noise. Coherent peaks at two-way travel times may then identify the ice--ocean and ocean--rock boundaries. Studies of Europa and terrestrial ice shelves show the potential of this method while also emphasizing the need for low incoherent noise and long, stable seismic records \citep{PanningEtAl2018,StahlerEtAl2018,ZhanEtAl2014,GomezGarciaEtAl2023}.

The feasibility of noise autocorrelation on Triton depends on the amplitude and spatial distribution of background seismic noise, which may arise from tectonic and impact activity, tidal deformation, and ocean--ice interaction, plus sensor self-noise and lander--ground coupling. Forward modeling using the PlanetProfile velocity structure remains necessary because autocorrelation-derived interface times suffer from an inherent trade-off between layer thickness and seismic wave speed. Complementing these seismic constraints, magnetic-induction measurements probe a distinct set of parameters: seismology primarily constrains interface depth and wave speed, whereas induction is sensitive to ocean electrical conductance. Joint interpretation of the two data types can therefore disentangle ocean thickness from conductivity more effectively than either method alone.

\endgroup

\section{Discussion and Recommendations for Investigating the Space Environment and Interior Structure of Icy Moon}

Understanding the internal structure of icy moons critically relies on magnetic field and current density measurements within their space environments. Detecting the weak induced magnetic fields poses a significant technical challenge. For instance, Triton's induced magnetic field intensity can be less than 10\% of the background magnetic field beyond 1000 km, with quasi-static magnetic fields on the order of 0.1 nT. This requires magnetometers to be optimally designed for flyby altitudes, ensuring high-precision measurements during close encounters, such as at altitudes below 100 km, as close as possible to Triton’s surface.

For current measurements, comprehensive coverage of the broad energy range (eV to keV) of plasma in planetary magnetospheres (like Neptune's) and icy moon ionospheres (like Triton's) is essential. We recommend employing a suite of Faraday cup retarding potential analyzers (RPAs). An optimal configuration would include two independent units, each with two probes, enabling omnidirectional, dual-energy, and multi-species particle detection. This setup is crucial for resolving the complex influence of space electric currents on magnetic field distributions. For example, space electric currents near Triton can induce magnetic fields stronger than the background field, significantly altering the spatial morphology of the magnetic field. Faraday cup RPAs have a proven track record as critical payloads for measuring plasma in planetary space environments, as exemplified by the Plasma Science (PLS) instruments on Voyagers 1 and 2, which conducted measurements of magnetospheric plasmas during encounters with Jupiter, Saturn, Uranus, and Neptune \cite{belcher1989plasma}. More recently, two identical Faraday cup systems (Plasma Instrument for Magnetic Sounding, PIMS) aboard the Europa Clipper mission, launched in October 2024, aim to explore the plasma environments in Jupiter’s magnetosphere and Europa’s ionosphere\citep{westlake2023plasma}.

Furthermore, space current detection payloads can be integrated with spectral imaging techniques to facilitate a "first observe, then probe" progressive survey of icy moon plumes. A pertinent example is the Cassini spacecraft's flyby of Enceladus, during which spectral imaging initially identified the plume distributions. Subsequently, in-situ particle detectors analyzed the charged particle characteristics of water, carbon dioxide, and other components, thereby establishing a material link between plume eruptions and the internal ocean\citep{waite2006cassini}.

\rev{Seismological constraints are evaluated through end‐to‐end synthetic‐waveform experiments. Three‐dimensional wave propagation in a layered Triton interior is modeled with ``SPECFEM3D\_GLOBE'', and a hierarchical travel‐time inversion recovers source parameters,ice‐shell thickness, and ocean thickness from direct and reflected P‐wave arrivals. A five‐station synthetic recovery test demonstrates that interface depths are resolvable to first order, with residual errors arising primarily from limited‐aperture depth sensitivity and velocity–thickness trade‐offs. Detection thresholds are derived from flight‐heritage VBB noise levels, scaled by a conservative environmental margin, and combined with amplitude‐scaling relations to estimate minimum detectable magnitudes as a function of distance. A multi‐station network improves source location and phase association, whereas a single well‐coupled station can exploit known impacts, repeating events, polarization, receiver‐side reverberations, and ambient‐noise autocorrelation under favorable source and noise conditions. Because Triton’s seismic event rate remains poorly constrained, long‐duration operation and sensitivity studies spanning a range of source strengths are warranted.}

\rev{The three principal measurement classes supply complementary constraints. Magnetic induction is sensitive to the combined conductivity and geometry of the ocean; in this context, Cochrane et al. (2022)~\cite{cochrane2022single} evaluated single- and multiple-flyby magnetometric detection of a Triton ocean, systematically addressing ionospheric contributions, external-field variability, trajectory uncertainty, data gaps, and observational noise. The present framework extends that observational logic by adding a multi-fluid calculation of plasma-current perturbations, which characterize magnetic contributions generated outside the interior, and a seismic constraint on interface depths.  PlanetProfile connects these measurements by requiring the electrical and seismic profiles to correspond to a common pressure--temperature--composition structure.}

\begin{figure}[htbp]%
\centering
\includegraphics[width=1.0\textwidth]{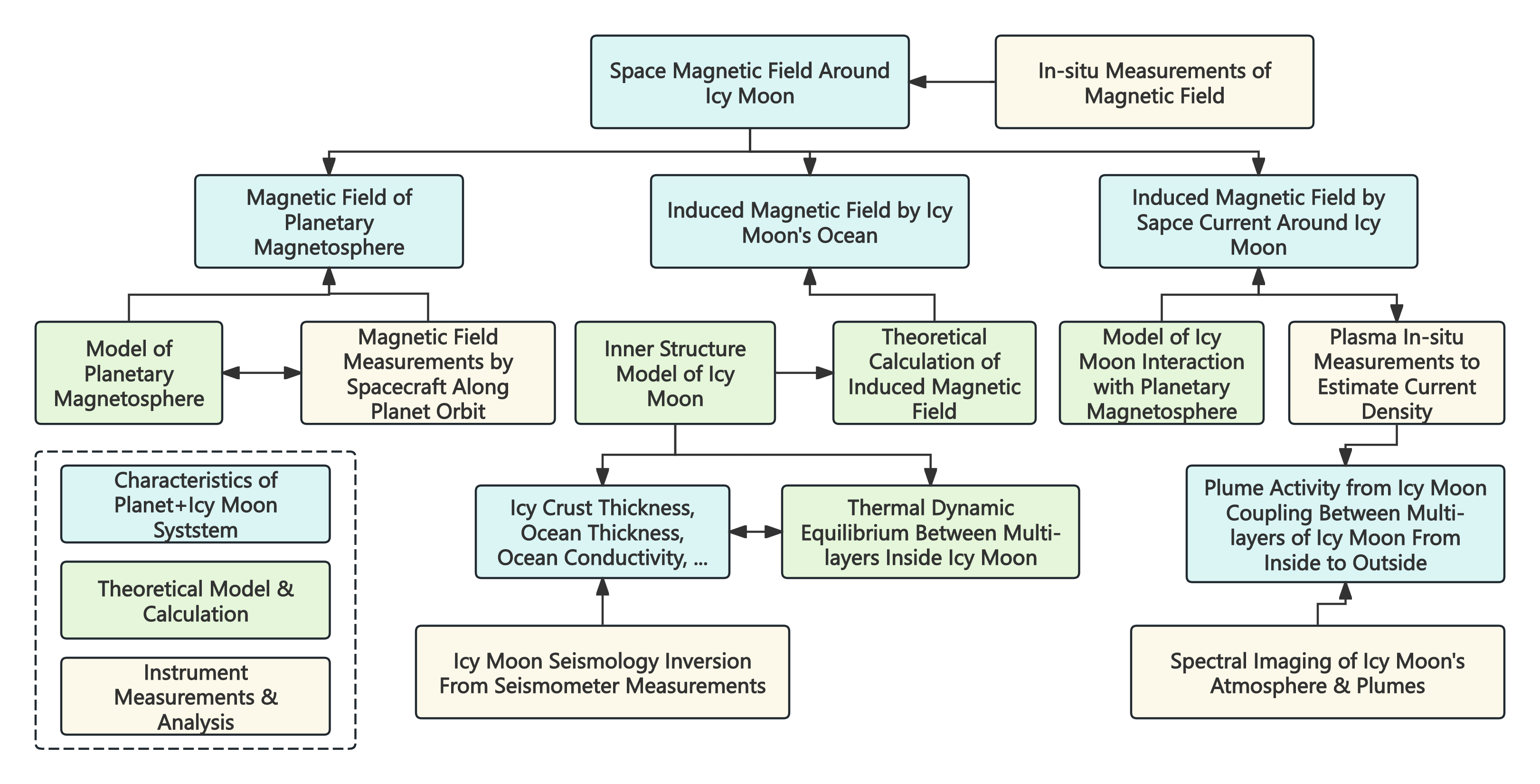}
\caption{\rev{Conceptual relationship between model components, observable quantities, and possible future measurements for interpreting icy-moon interiors and space environments. Cyan boxes denote target physical quantities or coupled-system properties, green boxes denote theoretical modeling and forward-calculation components, and beige boxes denote observational inputs or measurement strategies. The present study focuses on the central chain linking an interior-structure model, magnetic-induction calculations, plasma-interaction modeling, and seismic travel-time constraints on ice-shell and ocean thickness. Other elements shown in the diagram, such as spectral imaging of plumes and mission-specific magnetic, plasma, or seismic measurement strategies, indicate possible future extensions rather than calculations completed in this work.}
}\label{fig12}
\end{figure}

\rev{Figure~\ref{fig12} summarizes these model--observable connections conceptually and also marks several measurement strategies that would be needed in future mission-specific studies.}

\rev{Although not modeled here, gravity and tidal observations would provide critical independent constraints on the radial mass distribution, ice-shell rigidity, and ocean-shell mechanical decoupling. These measurements would help anchor the PlanetProfile reference model against which the magnetic-induction and seismic interpretations are evaluated. Radio-science Doppler tracking can constrain static low-degree gravity coefficients (\(J_2\), \(C_{22}\)) and the time-varying tidal Love number \(k_2\) through orbital perturbations. Surface-deformation measurements, including libration-amplitude monitoring from imaging and tidal-height determination from laser altimetry, can provide complementary constraints on shell-ocean coupling and the displacement Love number \(h_2\). We note that PlanetProfile's existing interface with pyalma3 makes complex \(k_2\) and \(h_2\) calculations technically feasible; however, such calculations require additional assumptions for ice and rock rheology, viscosity profiles, forcing periods, and thermal structure that are currently poorly constrained for Triton. Rather than introduce unconstrained Love numbers, we identify gravity/tidal modeling---enabled by the PlanetProfile--pyalma3 infrastructure---as a concrete future extension of the present framework \citep{VanceEtAl2018,marusiak2021exploration}.}

%%【待续，2025-07-27 10:03 pm】

\section{Summary}

\rev{This study demonstrates how coupled geophysical and space-physics modeling can identify complementary observables for probing Triton's ocean and surrounding plasma environment.} To implement this conceptual framework and establish a robust research paradigm, our study has undertaken \rev{four} core tasks: modeling Triton's internal layered structure; simulating the induced magnetic fields generated by conductive internal layers; investigating the interactions between Triton's neutral atmosphere, ionosphere, and Neptune's magnetosphere; \rev{and examining seismic-wave propagation and travel-time constraints on the ice shell and ocean}. \rev{Together, these calculations show how internal induction, plasma-current perturbations, and seismic interface constraints can be combined to interpret future Triton observations.}

For the adopted 10~g~kg$^{-1}$ MgSO$_4$ reference model, Triton contains a 113.75 km ice shell overlying a 124.80 km ocean. The ocean floor occurs near 201.3~MPa, approximately 330.7~MPa below the MgSO$_4$ ice-V onset pressure, so the modeled hydrosphere remains in the liquid stability regime down to the ocean--rock interface. Composition and core sensitivity tests show that the assumed ocean composition has a stronger effect on the induction response than the inclusion of an iron core for the cases examined here.

The magnetic calculations show that induction is sensitive to the combined conductivity and geometry of the ocean, but ocean thickness and conductivity can produce partially degenerate responses. The plasma calculation further shows that magnetic perturbations generated by Triton's surrounding current system can be comparable to the internal induction signal. Thus, interpreting future magnetic-field measurements requires simultaneous characterization of both the internal induction response and the external plasma-current contribution.

The seismic calculation provides a complementary constraint on layer geometry. In the controlled five-station synthetic recovery test, the ice-shell and ocean thicknesses are recovered with errors of 8.4\% and 12.5\%, respectively. Under the adopted 100-times-broadband-noise sensitivity case, the minimum magnitude required for the largest arrival increases from approximately \(M_w=3.8\) at 100~km to \(M_w=4.6\) at 1000~km. These results are conditional on the assumed source, station geometry, attenuation, geometrical spreading, and noise model, and should not be interpreted as formal mission-performance predictions.

Overall, the results indicate that magnetic induction, plasma measurements, and seismic travel times provide complementary observables for separating ocean conductance from interface depth. Future work should extend this framework with trajectory-specific magnetic/plasma simulations, ensembles of seismic sources and station geometries, measured or physically modeled Triton noise environments, and gravity/tidal constraints on the radial mass distribution and shell rigidity. Such extensions would inform requirements for magnetometers, plasma instruments, and landed seismometers, but detailed payload design and flight qualification remain beyond the scope of the present study.

% \begin{acknowledgements}
\section*{Acknowledgements}
This work was supported by the Pre-research Project on Civil Aerospace Technologies D010301 funded by the China National Space Administration (CNSA). This work was also supported by the National Natural Science Foundation of China and the Key Research and Development Program of the Ministry of Science and Technology of China (Grant Nos. 2021YFA0718600, 42241118, 42174194, 42150105, 2022YFF0503800). We would like to specify the contributions of individual authors as follows: Jiansen He proposed the research idea and designed the specific research content and methodology; Chuanpeng Hou and Haoen Xie performed most of the simulation calculations and figure preparation. \rev{Jiansen He, Chuanpeng Hou, and Haoen Xie wrote the manuscript and are co-first authors of this article;} Jiansen He is the corresponding author. Jiaqi Li provided assistance with calculations and figure plotting related to seismic wave propagation detection. Tianhang Chen, Hong Zou, Xuzhi Zhou, Hui Li, Yan Li, Fuchuan Pang, Bingkun Yu, Hui Huang and Tong Wang contributed to the discussions and revisions of the manuscript. In this study, we utilized code libraries including PlanetProfile, MoonMag, and SWMF to conduct simulation research on the internal structure, internal induced magnetic fields, and space environments of icy moons. We acknowledge the use of SPECFEM3D\_GLOBE for numerical simulations of seismic-wave propagation in Triton’s interior. We also thank Prof. Qiugang Zong and Prof. Dali Kong for valuable discussions on the research direction of this work. Additionally, we acknowledge the organization of the Shuangqing Forum (No. 377: "Exploration of Extraterrestrial Life and Evolution of Habitable Environments") and the Xiangshan Science Conference (No. 767: "Frontier Scientific Issues and Key Technologies in Ice Giant Exploration"), which provided academic exchange platforms and opportunities for the continuous improvement and refinement of this work.
% \end{acknowledgements}

% \bibliography{references}{}
% \bibliographystyle{elsarticle-num}

\end{document}